\documentclass[pre,showpacs,showkeys,nofootinbib]{revtex4}
\usepackage{graphicx}
\usepackage{amsmath}
\usepackage{color}

\newcommand{\rcite}[1]{Ref.~\cite{#1}}
\newcommand{\rcites}[1]{Refs.~\cite{#1}}
\newcommand{\eq}[1]{Eq.~(\ref{#1})}

\newcommand{\pnas}{{Proc.\ Nat.\ Acad.\ Sci.}}
\newcommand{\epl}{{EPL}}

\newcommand{\be}{{\bf e}}
\newcommand{\bv}{{\bf v}}
\newcommand{\br}{{\bf r}}
\newcommand{\bk}{{\bf k}}
\newcommand{\bF}{{\bf F}}
\newcommand{\meanL}{{q}}
\newcommand{\jeansT}{{{\cal T}}}

\begin{document}

\title{Dynamics of colloidal particles with capillary interactions}

\author{Alvaro Dom{\'\i}nguez}
\email{dominguez@us.es}
\affiliation{F\'\i sica Te\'orica, Universidad de Sevilla, Apdo.~1065, 
  E--41080 Sevilla, Spain}
\author{Martin Oettel}
\affiliation{Institut f\"ur Physik, WA 331, Johannes Gutenberg--Universit{\"a}t Mainz, 
  D-55099 Mainz, Germany}
\author{S.~Dietrich}
\affiliation{Max--Planck--Institut f\"ur Metallforschung, 
  Heisenbergstr.\ 3, D-70569 Stuttgart, Germany}
\affiliation{Institut f\"ur Theoretische und Angewandte Physik, 
  Universit{\"a}t Stuttgart, Pfaffenwaldring 57,
  D--70569 Stuttgart, Germany}

\date{March 17, 2010}


%

\begin{abstract}

  We investigate the dynamics of colloids at a fluid interface driven
  by attractive capillary interactions.  At submillimeter length
  scales, the capillary attraction is formally analogous to
  two-dimensional gravity. In particular it is a non-integrable
  interaction and it can be actually relevant for collective phenomena
  in spite of its weakness at the level of the pair potential. We
  introduce a mean--field model for the dynamical evolution of the
  particle number density at the interface. For generic values of the
  physical parameters the homogeneous distribution is found to be
  unstable against large--scale clustering driven by the capillary
  attraction. 
  We also show that for the instability to be observable,
  the appropriate values for the relevant parameters (colloid radius,
  surface charge, external electric field, etc.)  are experimentally
  well accessible.
  Our analysis contributes to current studies of the structure and
  dynamics of systems governed by long--ranged interactions and points
  towards their experimental realizations via colloidal suspensions.
\end{abstract}

\maketitle

\section{Introduction}
\label{sec:intro}

In recent years the issue of structure formation by colloids at fluid
interfaces has been the subject of intense experimental and
theoretical research. If the colloidal particles are only partially
wetted by the coexisting fluids, for not too high concentrations a
two--dimensional (2D) colloid layer forms at the fluid--fluid
interface because the detachment energy of a particle from such an
interface is much larger than the thermal energy \cite{Pier80,ChPi03}.
It turns out that these systems provide an excellent test-bed for
fundamental issues (such as, e.g., 2D phase transitions \cite{ZLM99})
as well as interesting perspectives for a variety of applications
(such as, e.g., micropatterning and novel materials emerging from
particle self--assembly \cite{VeVe06}).

The colloidal particles forming a 2D layer at fluid interfaces
interact with each other in diverse ways. For the present purpose,
these interactions can be classified into three groups (see, e.g.,
\rcite{OeDi08}). (i) There is the ubiquitous van der Waals force,
complemented by double--layer electrostatic interactions. These are
described by the Derjaguin--Landau--Verwey--Overbeek (DLVO) model
(see, e.g., \rcite{KlLa03}) and are relevant only if the particles
have a chance to come sufficiently close, leading to coagulation.
(ii) One can find also a repulsive interaction of longer range. The
presence of unscreened charges on the particle surface exposed to a
nonpolar fluid phase generates an unscreened dipole--dipole repulsion
\cite{Hurd85,FDO07,DFO08}. If both fluids are dielectric, the same
kind of repulsion can be created by polarizing the particles with an
external electric field \cite{AuSi08}. If the particles are
paramagnetic, a repulsion arises between magnetic moments induced by
an external magnetic field \cite{ZLM99}. If one of the fluids is a
nematic phase, director deformations induced by the anchoring boundary
conditions at the particle surface can also lead to an effective
repulsion \cite{SCLN04,ODTD09}. This second group of interactions is
usually promoted in experiments with the aim to stabilize the colloids
against coagulation, in which case one can effectively neglect the
DLVO--type force, as we shall do in the following. (iii) Finally, upon
integrating out the interfacial degrees of freedom one obtains an
effective, capillary interaction due to the colloid--induced
deformation of the fluid interface (see, e.g.,
Refs.~\cite{KrNa00,Domi10}).
There are several ways how the particles can deform the interface.  A
capillary attraction arises if a vertical force is exerted on the
particles, e.g., due to buoyancy \cite{Nico49,CHW81,KPIN92}, an
external electric field \cite{AuSi08}, or the effect of a substrate if
the lower fluid is a film of finite thickness (see, e.g.,
\rcite{KrNa00}). If the particles are non-spherical or the wetting
properties of their surfaces are not homogeneous, the fluid interface
is deformed anisotropically and the ensuing capillary force is
attractive or repulsive depending on the relative orientation of the
particles
\cite{Luca92,SDJ00,BSR00,FoGa02,vSH05,DKNB05,LAZY05,LNO08,LoPo09,MFV09}.

The interest in capillary forces on the micrometer scale intensified
recently when it was proposed \cite{NBHD02} as an explanation for the
puzzling attraction observed
\cite{Onod85,GhEa97,RGI97,SDJ00,QMMH01,TAKK03,GoRu05,CTNF05,CTHN06}
between micrometer--sized particles which were supposed to exhibit
only the kind of repulsion discussed above as point (ii). However,
various theoretical studies \cite{FoWu04,ODD05a,WuFo05,ODD05b,DOD08}
have shown that under the prevailing experimental conditions the
capillary attraction between two \emph{spherical} micrometer--sized
particles is too weak to be able to explain the observed attraction.
The reported effect might even turn out to be just an artifact of the
experimental sample preparation \cite{FMMH04}.
Nevertheless, the capillary forces between charged particles still
remains a topic of current research interest (see, e.g.,
\rcite{BDCK09} for a recent study considering deviations from
sphericity).

These theoretical studies concern only the interaction between two
isolated particles. The capillary interaction in the submillimeter
range is formally analogous to unscreened 2D electrostatics (or 2D
Newtonian gravity) \cite{DOD08} and thus non-integrable in the sense
of equilibrium statistical mechanics: the energy of configurations
governed by this interaction is hyperextensive and the capillary
attraction could be the relevant driving force for collective, genuine
many--body phenomena in spite of its relative weakness at the
two--body level. One of the goals of the present study is the
investigation of this possibility for the kind of presently accessible
experimental setups. \citet{Perg09} has recently pointed out the
possible relevance of this many--body effect as an explanation of the
stability of the clusters observed in colloids at fluid interfaces.
We shall show, however, that this claim is actually unfounded.

The second goal of this work is to introduce a theoretical framework
which addresses the dynamical aspects of \textit{collective} evolution
under capillary attraction. The corresponding studies published in the
literature so far report on the motion of either a single particle
exposed to an externally created interfacial deformation or of few
particles, usually two, following their own capillary attraction.
Usually the equation of motion is solved in the overdamped
approximation in order to relate the velocity with the capillary
forces, thus providing a means of interpreting the experimentally
recorded particle position as a function of time (see, e.g.,
\rcite{PDDV95} and, more recently, \rcites{LAZY05,VvMM05,BDCK09}). In
\rcite{SiJo05} the full combined problem of particle motion and
hydrodynamic flows in the fluid phases is addressed numerically with
due account of the interfacial deformation. This approach is applied
to clusters of two to four millimeter--sized particles. On that scale,
the capillary attraction is screened and effectively very short ranged
because it decays exponentially beyond the capillary length, which is
typically of the order of millimeter.

Beyond this interest in the detailed motion of individual particles it
seems that little attention has been paid to the overall evolution of
a colloid monolayer driven by its own, self--consistently determined
capillary force field. In this respect we are only aware of
\rcite{VAKZ01}, where the clustering is studied experimentally as well
as by means of molecular--dynamics simulations and interpreted within
the framework of a certain effective kinetics of aggregation. In an
attempt to be as realistic as possible, this latter theoretical
approach takes into account many effects simultaneously (capillary
forces, DLVO interactions, solvation forces, fluid streaming by
temperature inhomogeneities). As a consequence in this analysis the
specific signature of the capillary interaction on the dynamics is
masked. Here we study the evolution of the coarse--grained
particle--density field, with the diffusion at the interface being
driven by the interparticle capillary attraction at scales below the
capillary length. This is addressed within the mean--field
approximation as being appropriate for a non--integrable interaction
at that range of scales. The problem has a formal resemblance with the
evolution of a self--gravitating 2D fluid, which allows us to predict
a phenomenology akin to the process of cosmological structure
formation.

The present study can also be viewed as a contribution to current
investigations of the structure and dynamics of systems governed by
long--ranged interactions, which can exhibit rather peculiar
properties \cite{CDR09,BGM10}. Our analysis points towards explicit
experimental realizations of such systems via colloidal suspensions.

In Sec.~\ref{sec:meanfield} the mean--field model is introduced.
After some qualitative considerations illustrating the overall
picture, the model is formulated in terms of a set of coupled
equations for the particle number density at the interface and the
mean--field interfacial deformation. In Sec.~\ref{sec:cluster} the
time evolution predicted by this model is analyzed within two
different approximations which facilitate an analytical solution of
the problem. We first investigate the \emph{linear} stability of a
homogeneous particle distribution and find a clustering instability,
which is the analogue of the so-called \emph{Jeans' instability}
studied in the astrophysical literature. Then we compute the
\emph{non-linear} evolution of a radially symmetric perturbation in
the limit of strong capillary attraction (the so-called
\emph{cold--collapse} approximation following the cosmological
terminology). In Sec.~\ref{sec:experiments} we analyze available
experimental setups and conclude that there is an accessible range of
parameters for which one can expect that the predicted phenomenology
is observable. In Sec.~\ref{sec:discussion} we discuss pertinent
experimental observations reported in the literature in the light of
our results as well as the precise relationship between our study and
the work by \citet{Perg09}. Sec.~\ref{sec:conclusion} summarizes our
conclusions.

\section{Mean--field approach}
\label{sec:meanfield}

\subsection{Qualitative considerations}
\label{sec:qualitative}

\begin{figure}[h]
  \centering\includegraphics[width=.5\textwidth]{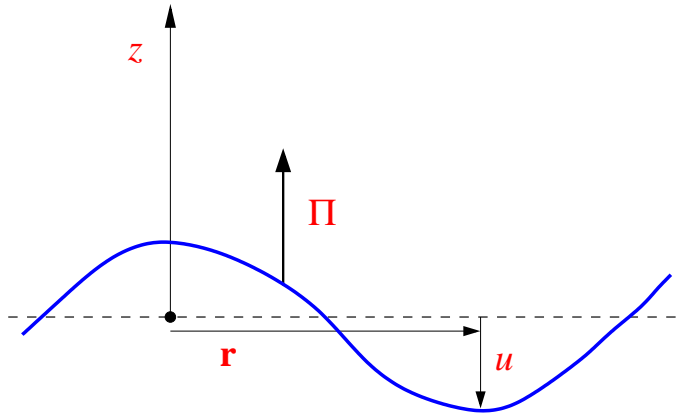}
  \caption{\textcolor{blue}{(color online)} Small deformation
    (exaggerated in this schematic drawing) of a fluid interface
    (solid line). $u(\br)$ is the vertical deformation field (i.e., in
    $z$~direction) with respect to the reference, flat configuration
    (dashed line) as function of the lateral coordinate $\br = (x,y)$.
    $\Pi(\br)$ is the vertical force per unit area acting on the
    interface in the reference configuration. These two fields are
    related by \eq{eq:laplaceu}.}
  \label{fig:smallDef}
\end{figure}

We recall briefly the electrostatic analogy of small interfacial
deformations, as worked out in \rcite{DOD08,Domi10} (see
Fig.~\ref{fig:smallDef}).
Let $u(\br)$ denote the small vertical deformation of an otherwise
planar interface at the lateral position $\br = (x,y)$, and $\Pi(\br)$
the vertical force per unit area exerted by external agents, e.g., a
pressure imbalance across the interface or forces directly exerted on
trapped particles (due to, e.g., buoyancy or optical tweezers).  These
two quantities are related by a Debye--H\"uckel--type equation,
\begin{equation}
  \label{eq:laplaceu}
  \nabla^2 u - \frac{u}{\lambda^2} = -\frac{1}{\gamma} \Pi ,
\end{equation}
where $\gamma$ is the surface tension and $\lambda$ is the capillary
length. This equation describes local mechanical equilibrium: at each
point of the interface, the force by external agents ($\Pi$) is
counterbalanced by the surface tension of the curved interface
($\nabla^2 u$), augmented by the force due to the weight of the fluid
displaced relative to the flat configuration ($u/\lambda^2$).
The lateral force on a piece $S$ of the interface exerted at its
border by the rest of the interface takes the simple
form 
\begin{equation}
  \label{eq:latF}
  \bF_{\rm lat} = \int_S d^2\br\; \Pi(\br) \nabla u(\br) .
\end{equation}
Equations~(\ref{eq:laplaceu}) and (\ref{eq:latF}) lead to the analogy
that the deformation $u$ plays the role of a 2D electrostatic
potential, $\Pi$ is the charge density, and the capillary
length\footnote{The capillary length is given by $\lambda =
  \sqrt{\gamma/(g |\bar{\rho}_1 - \bar{\rho}_2|)}$ in terms of the
  acceleration $g$ of gravity and the mass densities
  $\bar{\rho}_{1,2}$ of the coexisting fluid phases.} $\lambda$ is the
screening or Debye length, with the peculiarity that the forces are
reversed and charges of equal (different) sign attract (repel) each
other. Or, in other terms, the analogy holds with a ``screened''
gravitational interaction involving positive and ``negative'' masses.

\begin{figure}[h]
  \centering\includegraphics[width=.95\textwidth]{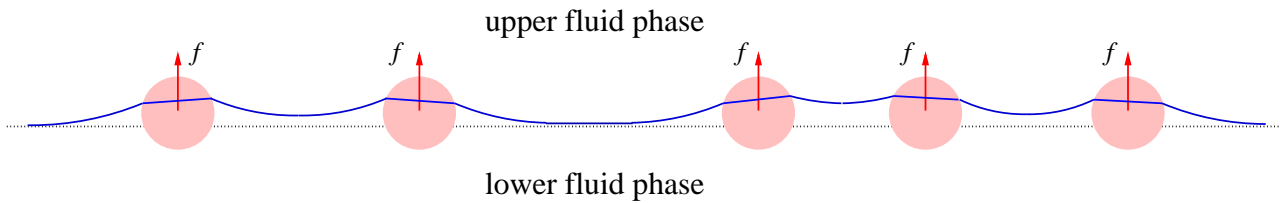}
  \caption{\textcolor{blue}{(color online)} Side view of a fluid
    interface containing many particles, each subjected to a vertical
    external force $f$. Within the simplest approximation the
    particles are modelled as pointlike capillary monopoles of
    strength $f$.
  }
  \label{fig:monopoles}
\end{figure}

In this respect, a particle trapped at the interface is characterized
by a set of (capillary) multipolar charges\footnote{If $R$ denotes the
  characteristic size of the particle, the multipolar expansion
  corresponds to, strictly speaking, an intermediate asymptotics at
  distances $R \ll r \ll \lambda$; for $r \gtrsim \lambda$, the
  expansion can be re-summed (see, e.g., \rcite{TBAW02}). The
  corrections to the multipolar expansion are, however, suppressed by
  powers of the small ratio $R/\lambda$. Since $R \sim 1$-$10\, \mu$m
  and $\lambda \sim 1\, \mathrm{mm}$ in the problems we are interested
  in, we can safely neglect this effect.} and the lateral force is
expressed in terms of the coupling of these multipoles with the
deformation field $u(\br)$. 
Thus, similarly, the capillary interaction between a collection of
particles (see Fig.~\ref{fig:monopoles}) can be written as the sum of
the interaction between pairs of multipoles localized at the position
of the particles.
In particular, the (isotropic) capillary monopole is simply equal to
the net vertical force $f$ exerted on the particle by external agents,
i.e., aside from the force provided by the interface
\cite{DOD08,DOD05}. The effective interaction between two monopoles of
equal strength $f$ separated by a lateral distance $d \lesssim
\lambda$ is described by the attractive potential
\cite{Nico49,CHW81,KPIN92,Domi10}
\begin{equation}
  \label{eq:Vcap_mono}
  V (d) \approx - \frac{f^2}{2\pi\gamma} \ln\frac{\lambda}{d} 
  \qquad (d < \lambda) .
\end{equation}
(Actually, the exact potential is given by $V(d) = -f^2
K_0(d/\lambda)/(2\pi\gamma)$ in terms of a modified Bessel function;
accordingly the length scale in the logarithm in \eq{eq:Vcap_mono} is
$\approx 1.12\lambda$ instead of $\lambda$, but we neglect this small
difference for the qualitative estimates to follow. At separations $d
\gtrsim \lambda$, the interaction is screened and the potential
crosses over to an exponential decay, which, for reasons of
simplicity, we set to zero in the following reasoning.)  This a 2D
Coulombic interaction, which is known to be non-integrable and to
drive an instability, e.g., in the astrophysically relevant case of a
self--gravitating gas. 

\begin{figure}[h]
  \centering\includegraphics[width=.5\textwidth]{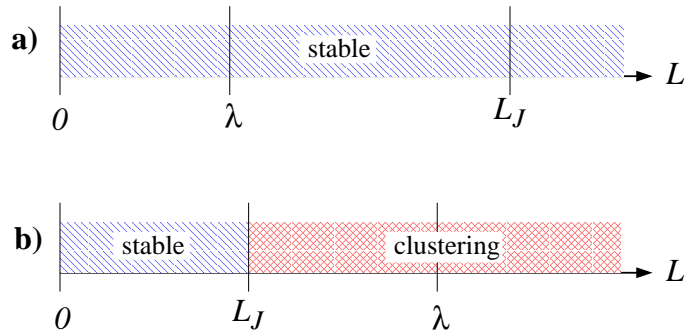}
  \caption{\textcolor{blue}{(color online)} Qualitative stability
    diagram of the homogeneous distribution in terms of the capillary
    length $\lambda$ and Jeans' length $L_J$ as a function of the
    system size $L$ for the two cases $\lambda < L_J$ (a) and $L_J <
    \lambda$ (b). }
  \label{fig:phases}
\end{figure}

At this point we consider it to be useful to introduce a qualitative
discussion which illuminates the physical origin of this instability.
The more formally interested reader can skip this part without loss in
favor of the mathematical derivation presented in the following
sections. Consider a collection of $N \gg 1$ identical monopoles at
positions $\br_i$ ($i=1,\dots, N$) distributed homogeneously over a
region of linear extension $L$ (and thus of average number density
$\varrho = N/L^2$). Due to the long--ranged nature of the interaction,
the capillary energy per particle $e_\mathrm{cap}$ of the
configuration can be estimated at the mean--field level as follows:
\begin{equation}
  \label{eq:ecap}
  e_\mathrm{cap} = \frac{1}{2 N} \sum_{i\neq j} V (|\br_i-\br_j|) 
  \simeq \left\{\begin{array}[c]{lc}
      \displaystyle \frac{1}{2} \int_{r<L} d^2\br\; 
      \varrho 
      \, V(r) 
      \sim - \frac{\varrho L^2 f^2}{8\gamma} 
      \left[ 1 + 2\ln\frac{\lambda}{L} \right] , & 
      \quad L < \lambda , \\
      & \\
      \displaystyle \frac{1}{2} 
      \int_{r<\lambda} d^2\br\; 
      \varrho 
      \, V(r) 
      \sim - \frac{\varrho \lambda^2 f^2}{8\gamma} ,
      & \quad \lambda < L .
    \end{array}\right.
\end{equation}
The key point is that for $L < \lambda$ this energy is {\rm not} an
intensive quantity but scales instead like $N = \varrho L^2$. There is
also a contribution $e_{\rm short}$ to the energy per particle, due to
thermal motion and short--ranged, predominately repulsive forces,
which is $N$--independent. Therefore, if the system is large enough
(but still $L < \lambda$) it can happen that the absolute value of the
energy due to capillary effects $e_\mathrm{cap}$ ($< 0$) dominates
over $e_\mathrm{short}$ ($> 0$).  The corresponding critical system
size can be characterized by \textit{Jeans' length} $L_J$ (this
terminology is borrowed from the astrophysical literature), which is
determined roughly by the condition $e_\mathrm{short} \simeq
|e_\mathrm{cap}|$:
\begin{equation}
  \label{eq:Lcrit}
  L_J \simeq \frac{1}{f} 
  \sqrt{\frac{8\gamma e_{\rm short}}{\varrho}} ,
\end{equation}
upon neglecting the logarithmic correction in \eq{eq:ecap}. (A precise
definition will be given in Sec.~\ref{sec:cluster}, see, c.f.,
\eq{eq:JeansK}.)  The capillary energy~(\eq{eq:ecap}) can be expressed
in terms of this length as follows:
\begin{equation}
  \label{eq:newecap}
  e_\mathrm{cap} \simeq 
  - e_\mathrm{short} \left(\frac{\lambda}{L_J}\right)^2 
  \times
  \left\{
    \begin{array}[c]{cc}
      \displaystyle \left(\frac{L}{\lambda}\right)^2 
      \left[1 + 2 \ln \frac{L}{\lambda} \right] , & L < \lambda , \\
      & \\
      1 , & \lambda < L .
    \end{array}\right.
\end{equation}
One can distinguish two distinct extremal cases (see
Fig.~\ref{fig:phases}):
\begin{itemize}
\item $\lambda < L_J$: One can easily see that in this case
  $|e_\mathrm{cap}| < e_\mathrm{short}$, independently of the lateral
  extension $L$ of the homogeneous particle distribution and the
  effect of the capillary attraction is just a small perturbation.
  This case corresponds to the physical situation in which the
  capillary force is screened (i.e., negligible at distances beyond
  the capillary length $\lambda$) preempting that its cumulative
  effect becomes comparable with the net effect of the short--range
  forces associated with $e_\mathrm{short}$.
\item $L_J < \lambda$: Here, $|e_\mathrm{cap}| > e_\mathrm{short}$
  whenever $L_J < L$, implying that the homogeneous distribution is
  unstable because the attractive capillary force cannot be
  counterbalanced by the nonzero compressibility provided by
  $e_\mathrm{short}$. The system collapses into an inhomogeneous,
  clustered state with a new typical size $L^*$ and a new average
  density $\varrho^*$ sustained by its own capillary attraction energy
  $e_\mathrm{cap}^*$, and for which the compressibility provided by
  $e_\mathrm{short}^*$ can balance the capillary attraction so that
  $|e_\mathrm{cap}^*| \simeq e_\mathrm{short}^*$ (and the new,
  effective Jeans' length $L_J^*$ is given by \eq{eq:Lcrit}). If
  Eq.~(\ref{eq:newecap}) is evaluated for the quantities carrying an
  asterisk (characterizing the clustered state), this condition has
  two possible solutions:
 \begin{itemize}
 \item[(i)] the cluster size is of the order of its effective Jeans'
   length, and the latter in turn is smaller than the capillary
   length, i.e., $L^* \sim L_J^* < \lambda$,
 \item[(ii)] the effective Jeans' length is of the order of the
   capillary length, and the latter in turn is smaller than the
   cluster size, i.e., $L_J^* \sim \lambda < L^*$.
  \end{itemize}
  (In Appendix~\ref{sec:collapse} we present an alternative
  qualitative derivation of condition (i) based on a force--balance
  argument rather than this energy consideration.)
\end{itemize}
These qualitative considerations are formalized in the following
sections using a mean--field model for the particle dynamics under the
action of capillary attraction.

\subsection{Mean--field capillary force}

We consider a system of particles trapped at a fluid interface
considered to be flat for the time being. In the absence of
long--ranged interactions, there are equilibrium states corresponding
to a 2D homogeneous fluid phase with areal particle number density
$\varrho$, characterized by an equation of state $p(\varrho)$, where
the ``pressure'' $p$ is the lateral force per unit length exerted by
the particles on the walls providing the lateral confinement of the
monolayer. (In what follows, we omit the possible dependence on the
temperature $T$ which is irrelevant in the present context because we
assume isothermal conditions maintained by the fluids on both sides of
the interface.) One can write (setting the Boltzmann constant equal to
unity)
\begin{equation}
  \label{eq:eqstate}
  p(\varrho) = \varrho T + p_{\rm ex} (\varrho) ,
\end{equation}
where the first term is the entropic (ideal gas) contribution and the
second term (excess pressure) is determined by the short--ranged
interactions between the particles. There are always the hard--sphere
contribution and the van der Waals attraction which eventually leads
to colloid coagulation. However, in the experiments aimed at observing
capillary effects these contributions are irrelevant because a
longer--ranged repulsion (of electric, magnetic, or elastic origin) is
implemented precisely to avoid coagulation \cite{OeDi08}. Typically
this repulsive potential has the simple form
\begin{equation}
  \label{eq:urep}
  v_{\rm rep} (d) = \frac{A}{d^n} = 
  T \left(\frac{\zeta}{d}\right)^n , \qquad (n>2),
\end{equation}
where the constant potential parameter $A$ can be replaced by the
Bjerrum length $\zeta = (A/T)^{1/n}$ of the potential (i.e.,
$v_\mathrm{rep} (d=\zeta) = T$, so that $\zeta$ decreases with
increasing $T$). The repulsion of electric or magnetic origin
corresponds to $n=3$ \cite{OeDi08,DFO08}, while the value of $\zeta$
can be varied via the externally controllable parameters of the system
(strength of the external magnetic or electric field, water salinity,
temperature, etc.). Because of the simple scaling behavior of this
potential, the phase diagrams of such kind of fluids are determined by
the single dimensionless parameter $\zeta^2 \varrho$ (see, e.g.,
\rcite{HGJ71}): for sufficiently small $\zeta^2 \varrho$ (high
temperature, i.e., small $\zeta$, or low density) there is a fluid
phase; if $\zeta^2 \varrho$ is above a certain value, depending on
$n$, the system freezes (for $n=3$ this threshold value is $\zeta^2
\varrho \approx 4.6$ as obtained from experimental data \cite{ZLM99}).
Since the system is two--dimensional, the fluid--solid transition is
expected to be of the Kosterlitz--Thouless type. For $n=3$ this is
supported experimentally \cite{ZLM99} and by recent simulations
\cite{LZT06}, according to which there is a narrow range of values of
the parameter $\zeta^2 \varrho$ within which a hexatic phase is
observed between the fluid and the solid phase.

In order to address the effect of the long--ranged capillary force,
one introduces the ensemble--averaged interfacial deformation
$U(\br)$. From \eq{eq:laplaceu} one obtains
\begin{equation}
  \label{eq:meanU}
  \nabla^2 U - \frac{U}{\lambda^2} = -\frac{f}{\gamma} \varrho ,
\end{equation}
in terms of the particle number density field $\varrho(\br)$. Here,
the ensemble--averaged vertical pressure $\Pi(\br)$ has been replaced
by the density of capillary monopoles,
\begin{equation}
  \Pi(\br) = \Pi_{\rm mon}(\br) = f\varrho(\br) ,
\end{equation}
where the capillary monopole $f$ associated with a
single particle is the net vertical force exerted on it by external
agents. 
By analogy with 2D electrostatics and gravity, one can apply a
mean--field approximation: the lateral capillary force experienced by
a single particle located at $\br$ is written as $+f\nabla U(\br)$,
after replacing the ensemble--average of $\Pi \nabla u$ in
\eq{eq:latF} by $f \varrho \nabla U$. Within this approximation
correlations are neglected because $U(\br)$ is computed from the field
$\varrho(\br)$ via \eq{eq:meanU}, rather than from the density field
{\it conditional} to the presence of a particle at $\br$. Therefore
$U(\br)$ is actually the coarse-grained correlate of the interfacial
deformation $u(\br)$, neglecting small--scale spatial variations.  The
physical assumption underlying the mean--field approximation is that
the dynamics of a particle is predominantly determined by the
simultaneous interaction with many other particles.  This is usually
expressed in terms of the constraint that the parameter $\varrho
\lambda^2$ ($\sim$ number of neighbors which de facto exert a force on
a particle) must be large.  In the experiments of interest here this
is always the case, because $\lambda \sim 1$ mm but the mean
interparticle separation ($\sim \varrho^{-1/2}$) lies in the
micrometer range.  (Another implicit assumption, peculiar to the
capillary problem, is that the capillary monopole $f$ of a particle is
independent of the presence of other particles, i.e., of the particle
density $\varrho$.  This is a good approximation if the vertical force
$f$ is predominantly due to gravity or due to an external electric or
magnetic field, see, e.g., \rcite{OeDi08} and references therein.)

In summary, the equilibrium state of the system is described
macroscopically by the force balance equation
\begin{equation}
  \label{eq:equil}
  -\nabla p(\varrho) + f \varrho \nabla U = {\bf 0} ,
\end{equation}
which, together with \eq{eq:eqstate} and \eq{eq:meanU}, determines the
equilibrium density profile $\varrho(\br)$. (Note that \eq{eq:equil}
also follows from, c.f., \eq{eq:diff}, or more generally
\eq{eq:relaxDyn}, for $\partial\varrho/\partial t = 0$.) Except for
the finite value of the parameter $\lambda$ in \eq{eq:meanU}, this
problem is formally analogous to the Vlasov--Poisson model
(corresponding to $\lambda=\infty$) for determining equilibrium
configurations of a fluid under its own self--gravity \cite{BiTr08}
(see \rcite{SiCh02} for a comprehensive analysis of the equilibrium
solution of the Vlasov--Poisson model for arbitrary spatial dimensions
with $p_\mathrm{ex}(\varrho)\equiv 0$, i.e., for an ideal gas, see
\eq{eq:eqstate}). In recent years this old problem has received
renewed attention and generalizations of it have been studied
throughly (see \rcite{SiCh10} and references therein for a brief
summary).

\subsection{Diffusive dynamics}

If the force balance equation (\ref{eq:equil}) is violated, the
particle density $\varrho(\br, t)$ will evolve in time according to
the law of mass conservation, expressed by the continuity equation
\begin{equation}
  \label{eq:cont}
  \frac{\partial \varrho}{\partial t} = -\nabla\cdot(\varrho \bv) .
\end{equation}
The flow velocity field $\bv(\br, t)$ is determined by the law of
motion of the particles. Each particle is dragged at the fluid
interface by a mean force $-(\nabla p)/\varrho + f \nabla U$. We
assume that the characteristic time scale of macroscopic evolution is
long enough so that the motion of the particles at the interface
occurs within the overdamped regime.  This allows one to neglect
particle inertia and a flow velocity is induced given by
\begin{equation}
  \label{eq:vel}
  \bv = \Gamma \left( - \frac{\nabla p}{\varrho} + 
    f \nabla U \right) ,
\end{equation}
where $\Gamma$ is a mobility coefficient of the particles at the
interface. In addition there are hydrodynamic interactions between the
particles due to the fluid flow induced by the particle motion which
can affect the dynamical evolution \cite{ZMM97}. The effect of this
interaction could be incorporated through a density dependence of
$\Gamma$ (see, e.g., \rcite{Batc72} for sedimenting hard spheres in
bulk fluids) or, in a more explicit manner, by additional terms in the
diffusion equation (\ref{eq:diff}) below \cite{ReLo09}.  For our
purposes, however, we neglect this effect and consider $\Gamma$ as a
phenomenological input parameter which is taken to be spatially
constant for reasons of simplicity\footnote{This approximation holds
  in the dilute limit.  Two particles of size $R$ moving in a
  three--dimensional bulk fluid at a distance $d \gg R$ acquire a
  relative velocity correction of the order $(R/d)$ \cite{RoPr69}. On
  the other hand for a solution of sedimenting hard spheres of radius
  $R$ and bulk number density $\varrho_b$, the mobility $\Gamma$ is
  corrected by a factor $\approx 1 - (3 R)^3 \varrho_b$ to lowest
  order in $\varrho_b$ \cite{Batc72}. These results can serve as a
  first estimate of the effect of the hydrodynamic interactions.
  However, one should keep in mind that the computation of these
  interactions in the presence of a deformable interface is actually
  an open problem, which lies beyond the scope of the present study.}.
Inserting this flow field into \eq{eq:cont} one obtains
\begin{equation}
  \label{eq:diff}
  \frac{\partial \varrho}{\partial t} = 
  \Gamma \nabla\cdot[\nabla p(\varrho) - f \varrho \nabla U] .
\end{equation}
On the other hand, we assume that the evolution of the areal number
density profile $\varrho(\br, t)$ occurs on a time scale sufficiently
large so that deviations from local equilibrium and the presence of
capillary waves can be neglected.  Therefore, the equilibrium
relationships in Eqs.~(\ref{eq:eqstate}) and (\ref{eq:meanU}) hold
and, when combined with \eq{eq:diff}, a closed equation is obtained to
determine the shape and the evolution of the density distribution
$\varrho(\br, t)$, which thus follows a diffusive dynamics driven by
``self--gravity''.
As discussed in Appendix~\ref{sec:functional} the problem can be cast
in terms of a functional formulation.

\section{Clustering instability}
\label{sec:cluster}

\subsection{Linear stability of a homogeneous state}

\begin{figure}[h]
  \centering\includegraphics[width=.5\textwidth]{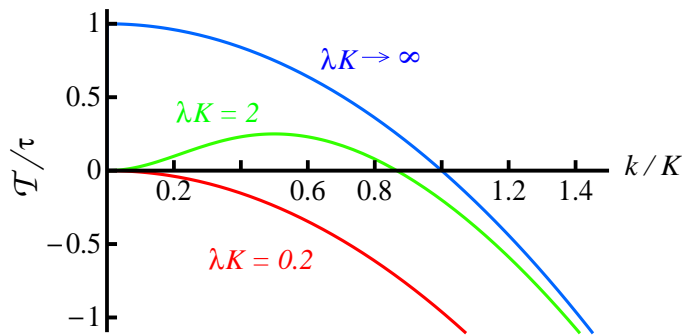}
  \caption{\textcolor{blue}{(color online)} Inverse relaxation time
    $1/\tau(k)$ (\eq{eq:disp}) as function of the wavenumber of the
    perturbation for different values of the parameter $\lambda K$.
    The units $\jeansT$ and $K$ are given by \eq{eq:JeansK}. For
    $\lambda K < 1$ one has $\tau(k) < 0$ and thus linear stability
    against perturbations of all wavenumbers. For $\lambda K > 1$
    perturbations with long wavelengths are linearly unstable due to
    $\tau(k) > 0$.}
  \label{fig:dispersion}
\end{figure}

\begin{figure}[h]
  \centering\includegraphics[width=.5\textwidth]{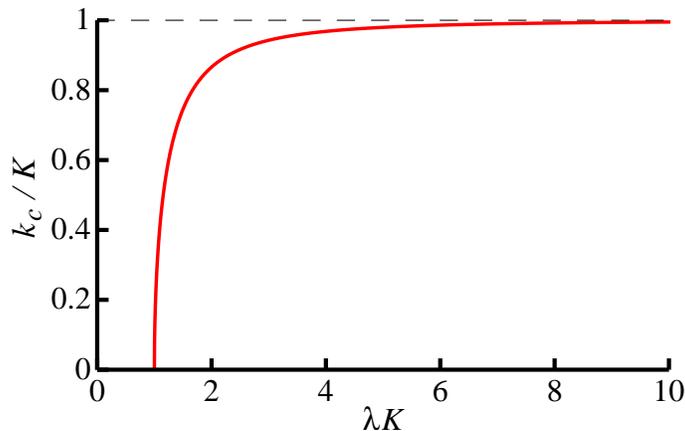}
  \caption{\textcolor{blue}{(color online)} Dependence of the critical
    wavenumber $k_c$ (\eq{eq:critk}) on the capillary length
    $\lambda$.}
  \label{fig:critk}
\end{figure}

The occurrence of a clustering instability can be inferred from
\eq{eq:diff}. For a macroscopically extended interface a homogeneous
particle distribution of density $\varrho_h$ is a solution of
Eqs.~(\ref{eq:meanU}) and (\ref{eq:equil}) with $U_h = f \lambda^2
\varrho_h /\gamma$ ( = const.). We consider now a perturbed
configuration $\varrho(\br,t) = \varrho_h + \delta\varrho (\br, t)$,
$U(\br,t) = U_h + \delta U (\br, t)$ and linearize
Eqs.~(\ref{eq:meanU}) and (\ref{eq:diff}) in terms of the small
perturbations $\delta\varrho$ and $\delta U$:
\begin{equation}
  \label{eq:linrho}
  \frac{\partial (\delta\varrho)}{\partial t} = \Gamma \nabla\cdot
  \left[ \frac{1}{\varrho_h \kappa_h}
    \nabla \delta\varrho - f \varrho_h \nabla \delta U 
  \right] ,
\end{equation}
\begin{equation}
  \label{eq:linU}
  \nabla^2 \delta U - \frac{\delta U}{\lambda^2} = -\frac{f}{\gamma}
  \delta\varrho ,
\end{equation}
where the isothermal compressibility $\kappa_h$ is given by
\begin{equation}
  \kappa_h := \left(\varrho 
    \frac{\partial p}{\partial\varrho}\right)^{-1}_T 
  (\varrho=\varrho_h) .
\end{equation}
We introduce the spatial Fourier decomposition of the perturbation,
\begin{equation}
  \widehat{\delta\varrho}(\bk, t) = \int d^2\br \;
  {\rm e}^{i\bk\cdot\br}
  \delta\varrho(\br, t) ,
\end{equation}
and define a characteristic wavenumber $K$ and a characteristic time
$\jeansT$ associated with the unperturbed homogeneous distribution
as
\begin{equation}
  \label{eq:JeansK}
  K^2 := \frac{f^2 \varrho_h^2 \kappa_h}{\gamma} , 
  \qquad
  \jeansT := 
  \frac{\gamma}{\Gamma f^2 \varrho_h} .
\end{equation}
The diffusion equation reduces to
\begin{equation}
  \frac{\partial \widehat{\delta\varrho}}{\partial t} = 
  \frac{1}{\tau (k)} \widehat{\delta\varrho} 
  \qquad\Rightarrow\qquad
  \widehat{\delta\varrho} (\bk, t) = 
  \widehat{\delta\varrho} (\bk, 0)\, {\rm e}^{t/\tau(k)} ,
\end{equation}
with a typical time of evolution $\tau(k)$ given by
\begin{equation}
  \label{eq:disp}
  \frac{1}{\tau (k)} = - \frac{\Gamma k^2}{\varrho_h \kappa_h} + 
  \frac{\Gamma f^2 \varrho_h}{\gamma}
  \frac{(\lambda k)^2}{1 + (\lambda k)^2} =
  \frac{1}{\jeansT} \left(\frac{k}{K}\right)^2 
  \left[ \frac{1}{(k/K)^2 + (\lambda K)^{-2}} - 1 \right] .
\end{equation}
In the astrophysical literature (in which $\lambda=\infty$ for
gravity, see, e.g., \rcites{BiTr08,Padm02} and references therein),
$K$ is known as Jeans' wavenumber (and $L_J := 1/K$ is the associated
Jeans' length, see \eq{eq:Lcrit}). The value of this parameter is
determined by the properties of the unperturbed homogeneous state.
Two qualitatively different cases can be distinguished (see
Fig.~\ref{fig:dispersion}):
\begin{itemize}
\item $\lambda K < 1$, so that $\tau (k) < 0$ for all values of $k$.
  In this case perturbations of all wavelengths decay exponentially as
  function of time; therefore the homogeneous state is stable.
  Physically, this describes the situation in which the number of
  particles inside the circle of interaction of radius $\lambda$ is
  too small and thus the capillary attraction is too weak to lead to a
  collapse of the colloidal fluid against a finite
  compressibility.
\item $\lambda K > 1$, so that $\tau (k) > 0$ for wavenumbers below a
  critical one,
  \begin{equation}
    \label{eq:critk}
    k_c = K \sqrt{1 - \frac{1}{(\lambda K)^2}} ,
  \end{equation}
  determined by the condition $\tau (k_c) = 0$. Perturbations
  satisfying this condition are linearly unstable, which describes the
  onset of a clustering instability.  In the limit of no screening of
  the capillary attraction (i.e., for $\lambda\to\infty$), one
  recovers the scenario of Jeans' instability: any homogeneous state
  is unstable against perturbations with a wavenumber smaller than
  Jeans' wavenumber. Figure~\ref{fig:critk} depicts $k_c$ as function
  of $\lambda$: for all practical purposes one can take $k_c \approx
  K$ unless the parameter $\lambda K$ is close to one. As can be
  inferred from Fig.~\ref{fig:dispersion}, $\jeansT$ is the
  characteristic time of the fastest growing mode if $\lambda K$ is
  not too close to one.
\end{itemize}


The value of Jeans' length $L_J = 1/K$ associated with every
homogeneous configuration determines its stability against clustering
by capillary attraction. If $L$ denotes the linear extension of the
system, the results are summarized by Fig.~\ref{fig:phases} obtained
previously in Subsec.~\ref{sec:qualitative} based on qualitative
arguments.  The dependence of stability on the equation of state of
each particular system enters only through the definition of Jeans'
length (\eq{eq:JeansK}).
A clustered phase is only possible if Jeans' length is small enough,
$\lambda K > 1$, {\em and} the system size is sufficiently large,
i.e., $2\pi < L K$, because a system of linear extension $L$ cannot
support perturbations with wavenumbers smaller than $2\pi/L$.
(Actually, if the system has a finite extension, the theoretical
analysis above has to be complemented with appropriate outer boundary
conditions compatible with a homogeneous particle distribution. We do
not expect this to change the conclusions qualitatively, but maybe the
precise value of $\lambda K$ setting the boundary of the stable region
in Fig.~\ref{fig:phases}.)

Finally, we remark that this Jeans--like instability has also been
analyzed in two recent studies. In \rcite{SiCh08} a model of
bacterial chemotaxis is considered, to which our equations reduce
formally in a certain limit. In \rcite{ChDe10} the same mathematical
model as ours is studied except that the system is confined to a
disk-shaped region with Neumann boundary conditions.

\subsection{Cold collapse of a radially symmetric perturbation}

Going beyond this simple linear analysis is possible only by resorting
to numerical computations. There is, however, a case which can be
addressed analytically: the dynamical evolution of a radially
symmetric perturbation of a homogeneous background in the limit of
vanishing Jeans' length, meaning physically an arbitrarily
compressible fluid (see \eq{eq:JeansK}). In the cosmological
literature this scenario is termed a ``cold collapse'' because in this
context one considers ideal gases; for them an infinite
compressibility amounts to a vanishing pressure $p(\varrho)$, which
corresponds to the limit of zero temperature.  The cold collapse is
therefore the limiting case of a more general scenario involving both
the capillary (gravitational) attraction and the pressure opposing
compression. This approximation allows one to obtain an exact
analytical solution of Eqs.~(\ref{eq:meanU}) and (\ref{eq:diff}) in
the presence of radial symmetry in the limit $\lambda\to\infty$.  The
computational details are presented in Appendix~\ref{sec:collapse}.
Here we just summarize the results: The evolution of a localized
radially symmetric perturbation of a homogeneous configuration is
driven predominantly by the capillary attraction provided the system
size is larger than Jeans' length; in this case collapse occurs only
if the spatially averaged density of the perturbation is larger than
the one of the homogeneous start configuration (i.e., if there is an
overdensity).  The collapse proceeds until a cluster size of the order
of the cluster Jeans' length (i.e., the one associated with the
density of the cluster) is reached and the pressure is able to halt
the collapse.  This last stage of the evolution, which does involve
the effect of a finite compressibility, is beyond the cold--collapse
approximation by construction and we have not studied it yet.
The total time of collapse is given by \eq{eq:Tcoll} within the
cold--collapse approximation and is roughly of the order of the
characteristic time $\jeansT$ associated with the homogeneous
configuration (see \eq{eq:JeansK}).

This exact solution of a simplified model can be used to gain some
insight into a more realistic situation. Suppose that in a homogeneous
configuration of density $\varrho_h$ Jeans' length is small enough so
that there can be found patches of linear extension $L < \lambda$
fulfilling the condition of instability, $L_J < L$ (see
Fig.~\ref{fig:phases}). In such a patch there are always thermally
induced density fluctuations $\delta \varrho$. Far from phase
transitions these exhibit a Gaussian distribution and are
uncorrelated. If such a patch is large enough to contain many
particles, the general theory of thermodynamic fluctuations states
that the relative amplitude of these fluctuations scales like the
inverse of the total number of particles in the patch, i.e.,
$|\delta\varrho|/\varrho_h \simeq 1/\sqrt{\varrho_h L^2} \ll 1$.  The
cold--collapse model can be employed to estimate the time of
\textit{coll}apse in that region with such an increased density (see
\eq{eq:Tcoll} with $\hat{\varrho}_0 = \varrho_h + |\delta\varrho|$):
\begin{equation}
  \label{eq:Tcoll2}
  \jeansT_{\rm coll} (L) \simeq 
  \jeansT \ln\sqrt{\varrho_h L^2} .
\end{equation}
Since $\jeansT_{\rm coll}(L)$ increases with $L$, one would expect a
{\it bottom--up} scenario of cluster formation in the language of
cosmology, according to which spatially smaller perturbations collapse
first, as opposed to a {\it top--down} scenario. However, the weak
logarithmic dependence implies that the bottom--up clustering would be
hardly observable; it is likely to observe the almost simultaneous
collapse of fluctuations of all sizes into a single cluster of maximal
size.

\section{Feasibility of experimental realizations}
\label{sec:experiments}

In the previous sections we have shown that the instability is
characterized by two parameters: Jeans' length $1/K$ and Jeans' time
$\jeansT$ (see \eq{eq:JeansK}). In this section we compute these
parameters for different setups which can be realized experimentally.
The clustering instability will be easily observable if one can find a
range of physical parameters for which the following constraints hold
simultaneously: (i) the particle size $R$ and Jeans' length $1/K$
should satisfy $R, 1/K < \lambda$ (Fig.~\ref{fig:phases}), (ii) the
mean interparticle separation $q$, which in this work will be measured
in units of the particle size $R$, i.e.,
\begin{equation}
  \label{eq:meanL}
  \meanL := \frac{1}{R\sqrt{\varrho}} ,
\end{equation}
should be small enough so that $\lambda/(\meanL R) \gg 1$ and the
mean--field predictions apply, and (iii) Jeans' time should lie within a
reasonable range which permits observations following up the collapse.
At this point we emphasize that in the present context the goal is to
observe a collective effect, i.e., a many--particle instability:
Capillary attraction is routinely observed between particles visible
for the naked eye ($R > 1\,\mathrm{mm}$), but in such a case the
capillary attraction is actually a force of very limited range
($\lesssim \lambda$) and the corresponding phenomenology is completely
different. Here, however, we focus our attention on micrometer--sized
particles so that $R \ll \lambda$.

The values of Jeans' length and Jeans' time depend on the specific
physical system via the strength $f$ of the capillary monopole and the
compressibility $\kappa$ determined by the equation of state
$p(\varrho)$ (see \eq{eq:JeansK}). We shall consider in detail three
different systems which are customarily employed in experiments: (i)
Charged particles or particles with dissociable surface groups, for
which $f$ is due to their weight and $p(\varrho)$ is determined by the
electrostatic interaction between the particles. (ii) Neutral
particles at the interface between two dielectric fluids in the
presence of an external electric field. In this case the capillary
monopole is due to both the weight of the particle and the electric
force exerted by the external field, while $p(\varrho)$ is again
determined by the electrostatic interparticle force. (iii)
Superparamagnetic particles in an external magnetic field. Here the
capillary monopole is due to their weight and the 2D equation of state
is determined by the magnetic interaction between the particles.  In
the following two subsections, we first compute Jeans' time, which
depends only on the monopole $f$, and then Jeans' length, which in
addition involves the equation of state.

In order to be specific, in the following we consider typical values
$\gamma = 0.07\, {\rm N/m}$ (surface tension of the air--water
interface at room temperature) and $1/\Gamma = 6\pi\eta_{\rm eff} R$,
where the effective viscosity $\eta_{\rm eff}$ interpolates between
the viscosities of the adjacent fluid phases (see, e.g., Fig.~4 in
\rcite{FDH06} and the measurements reported in \rcite{PDDV95}) and
depends on the contact angle at the particle--interface contact line;
we take $\eta_{\rm eff} = \eta_{\rm water}/2 = 0.5 \times 10^{-3}\,
{\rm Pa\times s}$ (corresponding to a sphere half immersed in water at
an air--water interface at room temperature). Because of the simple
scaling of Jeans' length and time with $\gamma$ and $\Gamma$ (see
\eq{eq:JeansK}) one can easily obtain the values of $1/K$ and
$\jeansT$ for other values of $\gamma$ and $\Gamma$ from the estimates
we shall quote below.

\subsection{Jeans' time}

\subsubsection{Monopole due to buoyancy}

We first compute Jeans' time $\jeansT$ (see \eq{eq:JeansK}). This time
depends only on the strength $f$ of the capillary monopole and is
independent of the detailed form of the interparticle repulsion. Every
particle at the interface carries a capillary monopole $f$ due to its
weight (corrected for the buoyancy effect due to the fluids).  A
spherical particle of radius $R$ floating at a fluid interface
experiences the vertical force
\begin{equation}
  \label{eq:fbuoy1}
  f_{\rm buoy} = \frac{4\pi}{3} g \bar\rho_{\rm eff} R^3 ,
\end{equation}
where $g$ is the acceleration of gravity and $\bar\rho_{\rm eff}$ is a
(signed) effective mass density, which depends on the mass densities
of the particle and of the fluids. For an estimate, we take as a
typical value $\bar\rho_{\rm eff} \approx \mbox{} - 1.6 \, {\rm g\times
  cm^{-3}}$ (corresponding to the glass particles at the interface
between air and corn oil used in the experiment described in
\rcite{ASJN08}), so that
\begin{equation}
  \label{eq:fbuoy2}
  f_{\rm buoy} \approx \mbox{} - \frac{16 T_\mathrm{room}}{\mu{\rm m}}
  \left(\frac{R}{\mu{\rm m}}\right)^3 ,
\end{equation}
with $T_\mathrm{room} = 300 K$ as the room temperature and the minus
sign indicating that the force points downwards.
Figure~\ref{fig:jeansT}(a) shows Jeans' time as a function of the
radius $R$ for several values of the average interparticle separation
$\meanL$. In this case, \eq{eq:JeansK} yields the simple scaling
${\cal T} \propto \meanL^2 R^{-3}$.

\subsubsection{Monopole due to external electric field}

\begin{figure}[h]
  \centering\includegraphics[width=\textwidth]{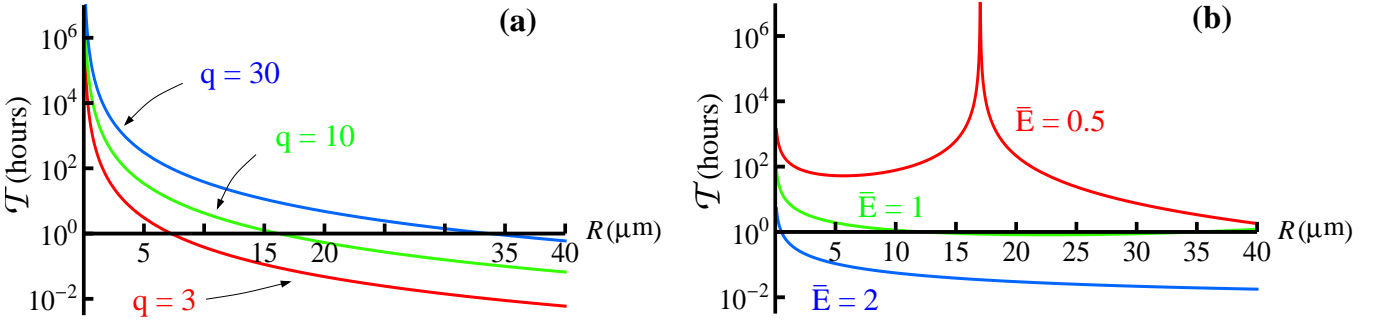}
  \caption{\textcolor{blue}{(color online)} Jeans' time for a
    homogeneous distribution of spherical particles as function of its
    radius $R$.  The particle mobility at the interface is taken to be
    (see main text) $\Gamma = (3\pi R)^{-1} \; {\rm m \times s/g}$ and
    the interfacial surface tension is chosen as $\gamma = 0.07\,{\rm
      N/m}$.  The 2D number density $\varrho_h = 1/(\meanL R)^2$ is
    parametrized in terms of $\meanL$, i.e., the mean interparticle
    separation divided by $R$, see \eq{eq:meanL}. In (a) the capillary
    monopole is given by \eq{eq:fbuoy2}, i.e., under the sole action
    of gravity, and corresponds to a vertical buoyancy force directed
    downwards.  In (b) the capillary monopole is given by
    \eq{eq:fTotal}, i.e., the particles are under the combined action
    of gravity and of an external electric field. The mean
    interparticle separation in units of $R$ is fixed at a value
    $\meanL = 30$ and the reduced electric field $\bar{E}$ is $E/
    (10^6\,{\rm V/m})$. The divergence of ${\cal T}$ at $R\approx
    15.7\,\mu\mathrm{m}$ for $\bar{E}=0.5$ corresponds to a vanishing
    total capillary monopole. For $\bar{E}=1$ and $\bar{E}=2$ this
    divergence occurs at $R\approx 63\,\mu\mathrm{m}$ and $R\approx
    252\,\mu\mathrm{m}$, respectively. For this reason the curve for
    $\bar{E}=1$ exhibits a minimum on the scale of the figure.}
  \label{fig:jeansT}
\end{figure}

Alternatively, a capillary monopole can be generated by a vertical
external electric field $E$ which polarizes the particles at the
interface between two dielectric fluids \cite{ASJN08,
  AuSi08}\footnote{The dipole induced in a particle creates an
  electric field $E_\mathrm{dip}(\br)$ decaying far from a particle
  asymptotically $\sim r^{-3}$. The action of this field on the
  interface induces an additional deformation not addressed in these
  studies. It can be computed from \eq{eq:laplaceu} with the electric
  pressure $\Pi \propto (E + E_\mathrm{dip}(\br))^2$, leading to an
  additional interfacial deformation $u \sim r^{-1}$. Here we neglect
  this contribution in the asymptotic comparison with the monopolar
  deformation, but acknowledge its potential relevance at high
  particle densities.}.  The vertical force on a spherical particle of
radius $R$ due to such a field $E$ is
\begin{equation}
  \label{eq:felec1}
  f_{\rm elec} = \varepsilon_0 R^2 E^2 (\epsilon_2 - \epsilon_1) 
  \phi ,
\end{equation}
where $\epsilon_1$ and $\epsilon_2$ are the dielectric constants of
the upper and lower fluid, respectively (if the electric field points
upwards) and the factor $\phi$ depends on the dielectric constants and
the height of the particle at the interface (see Fig.~7 in
\rcite{AuSi08}, where the factor $\phi$ is called $f_v$). With the
values $\epsilon_1 = 1$, $\epsilon_2 \approx 2.87$, $\phi \approx
0.27$, and $E\approx 10^6\,{\rm V/m}$ for the experiment described in
\rcite{ASJN08}, one obtains
\begin{equation}
  \label{eq:felec2}
  f_{\rm elec} \approx \frac{10^3 T_\mathrm{room}}{\mu{\rm m}} \left(
    \frac{E}{10^6\,{\rm V/m}} \frac{R}{\mu{\rm m}} \right)^2 ,
\end{equation}
and the total capillary monopole is the sum
\begin{equation}
  \label{eq:fTotal}
  f = f_{\rm buoy} + f_{\rm elec} \approx 
  \frac{T_\mathrm{room}}{\mu{\rm m}} \left(\frac{R}{\mu{\rm m}} \right)^2 \left[
    10^3 \left( \frac{E}{10^6\,{\rm V/m}} \right)^2 
    - \frac{16 R}{\mu{\rm m}} \right] .
\end{equation}
Figure~\ref{fig:jeansT}(b) shows Jeans' time as a function of the
radius $R$ for several values of the electric field for a fixed
particle density corresponding to a mean interparticle separation
$\meanL = 10$ (in units of $R$).
In this case, the electric field can compensate the weight and a
neutral buoyancy (i.e., $f=0$) can be achieved at a specific value
$R_*/\mu{\rm m} \approx [E/(1.26\times 10^5\, {\rm V/m})]^2$ of the
radius for a given electric field. Close to this value, Jeans' time
diverges as ${\cal T} \sim (R_* - R)^{-2}$. For radii much larger than
this critical value, ${\cal T}$ approximates the buoyancy dominated
regime discussed before, while in the limit $R \ll R_*$ the capillary
monopole is dominated by the electric force, i.e., $f \approx
f_\mathrm{elec}$ and ${\cal T} \propto \meanL^2 R^{-1} E^{-4}$.

\subsubsection{Monopole due to external magnetic field}
\label{sec:magneticMonopole}

In an experimental setup similar to the one just considered,
superparamagnetic particles can be placed in an external magnetic
field $H$ perpendicular to the interface which induces a capillary
monopole due to the magnetic vertical force which is described
analogously to \eq{eq:felec1}. However, the small values of the
magnetic susceptibilities $\chi_1$ and $\chi_2$ of the upper and lower
fluid phase, respectively (typically $|\chi_{1,2}| \sim 10^{-5}$)
renders this force irrelevant under usual experimental conditions.
Similar to \eq{eq:felec1} one can obtain the estimate (up to a
geometrical factor of order 1)
\begin{equation}
  f_{\rm mag} \sim \mu_0 R^2 H^2 
  \left(\frac{1}{1+\chi_2} - \frac{1}{1+\chi_1}\right)
  \approx \frac{10^{-5} T_\mathrm{room}}{\mu{\rm m}} \left(
    \frac{H}{10^2\,{\rm A/m}} \frac{R}{\mu{\rm m}} \right)^2 ,
\end{equation}
which, even for strong magnetic fields, is much smaller than the
capillary monopole due to weight given by \eq{eq:fbuoy2}. Therefore,
in the following we shall no longer consider the effect of a
magnetically induced monopole\footnote{In the experiments described in
  \rcites{ZMM97,ZLM99}, the particles are completely wetted by water
  and thus remain submerged, but very close to the interface.
  However, it is conceivable that superparamagnetic particles can be
  prepared which are only partially wetted by the fluid phases and
  thus get trapped at the interface.}.
This notwithstanding, in the following subsection we shall study the
influence of an external magnetic field on the compressibility of a
monolayer of superparamagnetic particles and thus on Jeans' length for
gravity induced capillary monopoles.

\subsection{Jeans' length}

\subsubsection{Compressibility for repulsive interactions}

The determination of Jeans' length requires the specification of the
equation of state of the 2D fluid of colloidal particles (see
Eqs.~(\ref{eq:eqstate}) and (\ref{eq:JeansK})). The simplest case is
that of an ideal gas, $p_{\rm ex} (\varrho) \equiv 0$; the
corresponding Jeans' length is given by
\begin{equation}
  \label{eq:Kideal}
  \frac{1}{K_\mathrm{ideal}} = \sqrt{\frac{\gamma T}{f^2 \varrho}} .
\end{equation}
Here the capillary monopole is only due to buoyancy so that $f =
f_\mathrm{buoy}$ (see \eq{eq:fbuoy2}) and $K_\mathrm{ideal}^{-1}
\propto q R^{-2}$.  This length is plotted in, c.f.,
Fig.~\ref{fig:Kcharged}(a) for reference. Similarly, one can consider
the reference case of a 2D gas of hard disks of radius $R$ (which
coincides with the radius of the spherical colloidal particles if
they are half--immersed in one of the fluids). The equation of state
is described well by the expression \cite{GZB97}
\begin{equation}
  \label{eq:eqStateHS}
  p = \varrho T 
  \frac{\varrho_c + \varrho}{\varrho_c - \varrho} ,
\end{equation}
where $\varrho_c := 1/(2\sqrt{3} R^2)$ is the number density for
close packing of disks in 2D. The corresponding Jeans' length can be
written as
\begin{equation}
  \label{eq:Kreal}
  \frac{1}{K} = \frac{{\cal L}(R^2 \varrho)}{K_\mathrm{ideal}} ,
\end{equation}
where $1/K_\mathrm{ideal}$ is Jeans' length of an ideal gas at the
same temperature and with the same number density (see
\eq{eq:Kideal}) and
\begin{equation}
  \label{eq:funcL}
  {\cal L} := \frac{1}{\sqrt{\varrho T \kappa}} =
  \sqrt{\frac{1}{T} \frac{d p}{d\varrho}}
\end{equation}
is a dimensionless function with ${\cal L}(0)=1$ which collects the
deviations from the ideal gas behavior. After taking \eq{eq:eqStateHS}
into account this correction of the ideal gas behavior is a function
${\cal L}(R^2 \varrho = q^{-2})$ solely of the dimensionless parameter
$\meanL$ defined in \eq{eq:meanL}, which must be larger than its value
$\meanL_c = 1/(R \sqrt{\varrho_c}) \approx 1.861$ at close packing.
One finds that this correction is significant actually only very close
to $\meanL=\meanL_c$; e.g., for $\meanL = 3$ one has ${\cal L} \approx
2.07$ so that in practical terms the curves in, c.f.,
Fig.~\ref{fig:Kcharged}(a) are applicable also for hard disks.

\begin{figure}[h]
  \centering\includegraphics[width=\textwidth]{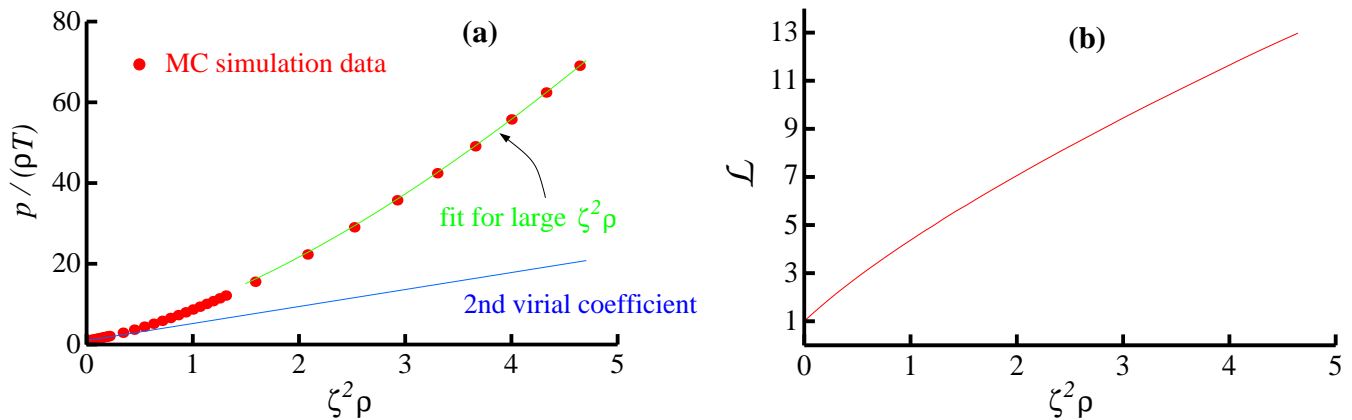}
  \caption{\textcolor{blue}{(color online)} 2D fluid of particles
    interacting according to the pair potential given by \eq{eq:urep}
    with $n=3$. (a) Pressure $p$ as a function of $\zeta^2 \varrho$,
    so that $p/(\varrho T) \to 1$ for $\varrho \to 0$.  The dots are
    the results from a MC simulation, while the lines correspond to
    the approximation of the equation of state up to the second term
    in the virial expansion, $p \approx \varrho T [1 + (\pi/2)
    \Gamma(1/3) \zeta^2 \varrho]$, valid for low densities, and to the
    numerical fit $p \approx \varrho T [3 + 6.6 (\zeta^2
    \varrho)^{3/2}]$, valid for large densities. Note that this latter
    fitting function corresponds to the equation of state of an
    harmonic solid of soft particles interacting with the pair
    potential $r^{-3}$ \cite{Spee03}. We have found that in good
    approximation this equation of state also holds for the
    corresponding high--density fluid phase. For $\zeta^2 \varrho
    \gtrsim 4.6$ the fluid freezes.  (b) The dimensionless function
    ${\cal L}$ given by \eq{eq:funcL} which characterizes Jeans'
    length (\eq{eq:Kreal}).}
  \label{fig:eqState}
\end{figure}

However, neither an ideal gas nor a gas of hard disks correspond to
the generic experimental situation. Typically the particles are
endowed with a soft interparticle repulsion in order to avoid
coagulation brought about by attractive dispersion forces. This kind
of repulsion is described by \eq{eq:urep}, where $n=3$ corresponds to
the experimentally relevant situations we want to address. As stated
in the context of \eq{eq:urep}, for such interaction potentials the
phase diagram depends on the parameter $\zeta^2 \varrho$ only. We have
used MC simulations (details can be found in Appendix~\ref{sec:MC}) to
compute the equation of state of this specific 2D fluid (see
Fig.~\ref{fig:eqState}(a)), for which we are not aware of published
data (compare \rcite{HGJ71} for other values of $n$ in a 3D fluid).
This equation of state is valid only for $\zeta^2 \varrho \lesssim
4.6$; beyond this value the 2D liquid freezes into a solid phase
\cite{ZLM99}.
Jeans' length can be expressed again as in \eq{eq:Kreal}, but the
function ${\cal L}$ defined by \eq{eq:funcL} is now a dimensionless
function of the parameter $\zeta^2 \varrho$ (see
Fig.~\ref{fig:eqState}(b)). At low densities there is a relatively
weak divergence, $1/K \propto 1/\sqrt{\varrho}$ (see \eq{eq:Kideal}),
reflecting the increase of Jeans' length caused by a weakening of the
overall capillary attraction upon dilution. At high densities close to
the onset of freezing Jeans' length increases slowly:
$1/K \sim \varrho^{1/4}$ because ${\cal L} \sim (\zeta^2
\varrho)^{3/4}$ as provided by a numerical fit and due to
$1/K_\mathrm{ideal} \sim 1/\sqrt{\varrho}$ (see \eq{eq:Kreal}).

\subsubsection{Repulsion between electrically charged particles}

\begin{figure}[h]
  \centering\includegraphics[width=\textwidth]{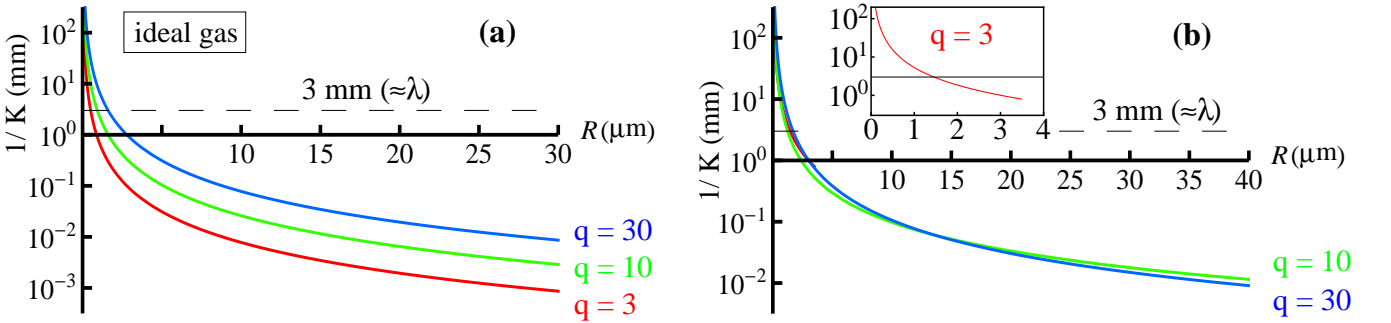}
  \caption{\textcolor{blue}{(color online)} Jeans' length for buoyancy
    induced monopoles (see \eq{eq:fbuoy2}) and (a) for the ideal gas
    equation of state (see \eq{eq:Kideal}) (also valid for a gas of
    hard disks, see the discussion after \eq{eq:funcL}), or (b) for
    the equation of state corresponding to dipolar repulsion between
    electrically charged particles estimated by taking charge
    renormalization into account (see Eqs.~(\ref{eq:Kreal}) and
    (\ref{eq:zetaCharged})).  $\meanL$ is the mean interparticle
    separation in units of $R$ (\eq{eq:meanL}).  The dashed,
    horizontal line indicates the value of $3\;$mm corresponding to a
    typical value of the capillary length $\lambda$.
    Capillary--induced clustering is possible below this line (see
    Fig.~\ref{fig:phases}). (The curve for $\meanL = 3$ in the inset
    stops at $R \approx 4\,\mu$m due to the onset of freezing.)}
  \label{fig:Kcharged}
\end{figure}

A common way of implementing the interparticle repulsion is to use
electrically charged particles or, more frequently, cover the particle
surfaces with chemical groups which dissociate in water. (Even then,
the repulsion is long--ranged if the adjacent fluid phase is a
dielectric \cite{DFO08}.)  The repulsion is described by \eq{eq:urep}
with $n=3$ and \cite{FDO07}
\begin{equation}
  \left(\frac{\zeta}{R}\right)^3 = 2 \frac{\epsilon_1}{\epsilon_2} 
  \frac{R}{\lambda_B} \left[ (1+\cos\theta)\, g_{\rm lin}\,
    \ln \left( 
    \frac{4\pi \sigma_c \lambda_B}{e\kappa_D}
  \right) 
  \right]^2 .
\end{equation}
Here, $1 < \epsilon_1 \lesssim 3$ and $\epsilon_2\approx 80$ are the
dielectric constants of air (or oil) and water, respectively,
$\lambda_B \approx 0.7\,{\rm nm}$ is the Bjerrum length in water at
room temperature, $\kappa_D^{-1} \lesssim 1\,\mu{\rm m}$ is the Debye
screening length in water, $e$ is the positive elementary charge,
$\sigma_{\rm c}$ is the charge density at the surfaces of the
particles in contact with water, $\theta$ is the contact angle at the
particle--interface contact line, and $0.1 \lesssim g_{\rm lin}
(\theta, \kappa_D) \lesssim 1$ is a factor of geometrical origin. For
strongly charged colloids one typically has $\sigma_c \sim 0.5\,
e/{\rm nm^2}$, while water with a salt concentration above
$10^{-2}\,$M has a screening length $\kappa_D^{-1}$ below a few
nanometers. Under these conditions, the factor $[(1+\cos\theta)\,
g_{\rm lin}\, \ln (4\pi \sigma_c \lambda_B/e\kappa_D) ]^2$ is of the
order of the unity, and in the following quantitative estimates we
replace it by 1.  Furthermore, we take $\epsilon_2/\epsilon_1 \approx
40$ as appropriate for an oil--water interface.  For these values of
the parameters one has
\begin{equation}
  \label{eq:zetaCharged}
  \zeta^2 \varrho \approx \frac{17.22}{\meanL^2} 
  \left(\frac{R}{\mu{\rm m}}\right)^{2/3} .
\end{equation}
For this expression Fig.~\ref{fig:Kcharged} shows Jeans' length $1/K$
(see \eq{eq:Kreal}) as a function of the particle radius $R$ for
various values $\meanL$ of the mean interparticle separation in units
of $R$.

\subsubsection{Repulsion between induced electric dipoles}

\begin{figure}[h]
  \centering\includegraphics[width=\textwidth]{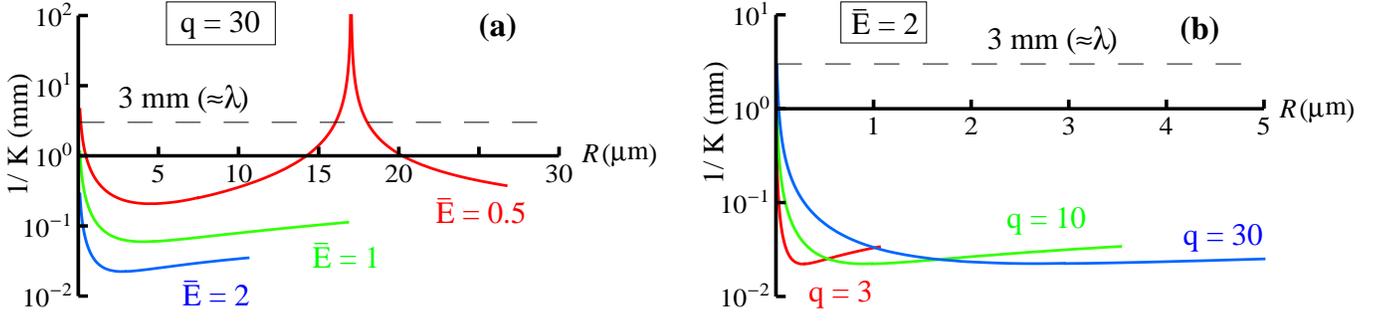}
  \caption{\textcolor{blue}{(color online)} Jeans' length $1/K$ for
    buoyancy and electric field induced monopoles (see \eq{eq:fTotal})
    and for the equation of state corresponding to electric field
    induced dipolar repulsion (see Eqs.~(\ref{eq:Kreal}) and
    (\ref{eq:zetaElectric})). In (a) Jeans' length is reported for
    various values $\bar{E}$ of the electric field in units of
    $10^6\,\mathrm{V/m}$ and for a fixed mean interparticle separation
    $\meanL = 30$ (in units of $R$).
    The divergence of $1/K$ at $R\approx 15.7\,\mu\mathrm{m}$ for
    $\bar{E}=0.5$ corresponds to a vanishing total capillary monopole.
    In (b) various values of $\meanL$
    for a fixed electric field $E=2\times 10^6\,\mathrm{V/m}$ are
    studied. The dashed, horizontal line indicates the value of
    $3\;$mm corresponding to a typical value of the capillary length
    $\lambda$. Capillary--induced clustering is possible below this
    line (see Fig.~\ref{fig:phases}). (When a curve stops, it does so
    at the corresponding onset of freezing.)}
  \label{fig:Kelectric}
\end{figure}

In the experiment described in \rcite{ASJN08}, the dipoles induced by
the external electric field give rise to an interparticle repulsion
described also by \eq{eq:urep} with $n=3$ and
\begin{equation}
  \left(\frac{\zeta}{R}\right)^3 = 
  - \frac{\varepsilon_0 R^3 E^2}{3 T} (\epsilon_2 + \epsilon_1) 
  \psi ,
\end{equation}
where the factor $\psi$ depends on the dielectric constants and the
height of the particle at the interface (see Fig.~9 in \rcite{AuSi08},
where the factor $\psi$ is called $f_d$). For the experiment in
\rcite{ASJN08} one has $\epsilon_1 = 1$, $\epsilon_2 \approx 2.87$,
and $\psi \approx - 0.019$ so that
\begin{equation}
  \label{eq:zetaElectric}
  \zeta^2 \varrho \approx \frac{14.00}{\meanL^2} 
  \left(\frac{E}{10^6\,{\rm V/m}}\right)^{4/3}
  \left(\frac{R}{\mu{\rm m}}\right)^{2} 
\end{equation}
at room temperature.  For this expression, Fig.~\ref{fig:Kelectric}
shows Jeans' length $1/K$ (see \eq{eq:Kreal}) as a function of the
particle radius $R$.


\subsubsection{Repulsion between induced magnetic dipoles}

\begin{figure}[h]
  \centering\includegraphics[width=\textwidth]{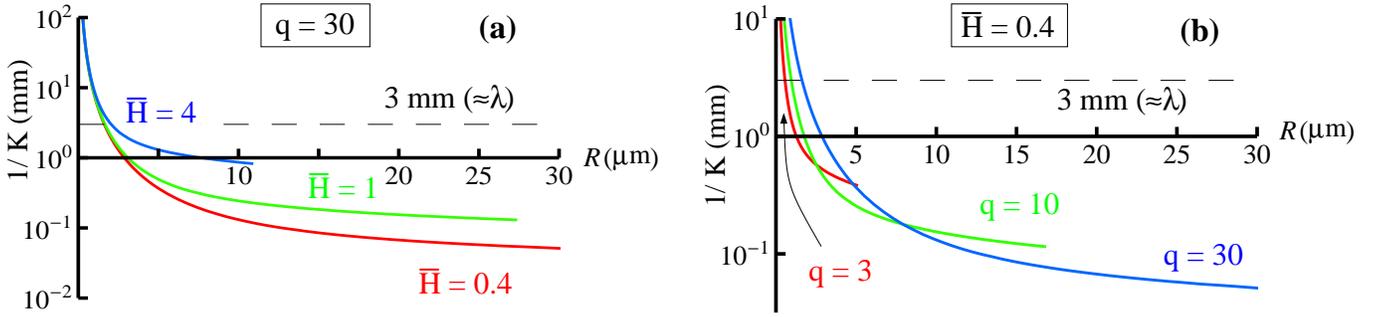}
  \caption{\textcolor{blue}{(color online)} Jeans' length $1/K$ for
    buoyancy induced monopoles (see \eq{eq:fbuoy2}) and for the
    equation of state corresponding to magnetic field induced dipolar
    repulsion (see Eqs.~(\ref{eq:Kreal}) and (\ref{eq:zetaMagnetic})).
    In (a) various values of the reduced magnetic field $\bar{H} = H /
    (10^2\,\mathrm{A/m})$ are studied for a fixed mean interparticle
    separation $\meanL = 30$ (in units of $R$).  In (b) Jeans' length
    is shown for various values of $\meanL$ 
    at the fixed field strength $H=40\,\mathrm{A/m}$ corresponding to
    Earth's magnetic field. The dashed, horizontal line indicates the
    value of $3\;$mm corresponding to a typical value of the capillary
    length $\lambda$. Capillary--induced clustering is possible below
    this line (see Fig.~\ref{fig:phases}). (When a curve stops, it
    does so at the corresponding onset of freezing.)}
  \label{fig:Kmagnetic}
\end{figure}

In the experimental setup with superparamagnetic particles of
susceptibility $\chi_\mathrm{m}$, an external magnetic field $H$
induces a magnetic moment, which in a spherical particle of radius $R$
is given by
\begin{equation}
  m = \frac{4 \pi}{3} R^3 \chi_\mathrm{m} H .
\end{equation}
In turn this leads to a dipolar repulsion described also by \eq{eq:urep}
with $n=3$ and
\begin{equation}
  \zeta^3 = \frac{\mu_0 m^2}{4\pi T} .
\end{equation}
In the experiments described in \rcites{ZMM97,ZLM99} the
susceptibility is $\chi_\mathrm{m} \approx 1.7$ and the magnetic field
$H$ ranges typically between $10^2\,$A/m and $10^3\,$A/m, which
corresponds to
\begin{equation}
  \label{eq:zetaMagnetic}
  \zeta^2 \varrho \approx \frac{5.31}{\meanL^2} 
  \left(\frac{H}{10^2\,{\rm A/m}}\right)^{4/3}
  \left(\frac{R}{\mu{\rm m}}\right)^{2} 
\end{equation}
at room temperature. (For comparison, Earth's magnetic field has a
strength of about $40\,$A/m and sets a lower bound on the value of $H$
achievable in the laboratory, unless the magnetic field is generated
in a specific configuration so as to counterbalance Earth's field.)
Figure~\ref{fig:Kmagnetic} shows Jeans' length $1/K$ (see
\eq{eq:Kreal}) as a function of the particle radius $R$. (As remarked
in Subsec.~\ref{sec:magneticMonopole}, the contribution to the
capillary monopole due to the magnetic field is negligible.)

\section{Discussion}
\label{sec:discussion}

Our results show that there is a range of parameters for presently
accessible experimental setups which allows one to observe the
instability of colloidal monolayers at fluid interfaces driven by
capillary attraction: particle size in the micrometer range, Jeans'
length well below the capillary length, and Jeans' time ranging from
minutes to weeks.
This provides theoretical evidence for the possibility that capillary
attraction can lead to relevant aggregation effects at submillimeter
length scales in spite of its relative weakness at the mean distances
between the particles.

In view of the estimates derived in Sec.~\ref{sec:experiments}, it is
conceivable that many experiments carried out so far happen to operate
in the range of parameters within which the homogeneous configuration
is stable against capillary--induced clustering. For
charge--stabilized colloids (see Fig.~\ref{fig:Kcharged}(b)) the
particles employed experimentally are usually too small (not larger
than a few micrometers). In the experiments carried out with
superparamagnetic particles \cite{ZLM99,ZMM97} (see
Fig.~\ref{fig:Kmagnetic}), the particles are too small ($R \approx
2.5\,\mu$m), the magnetic fields too large, and it is likely that the
capillary monopole is also too small (because the particles are
completely submerged in water, albeit close to the interface). In
\rcite{ASJN08}, which reports experiments in an electric field with
spherical particles ranging in radius between $2\,\mu\mathrm{m}$ and
$77\,\mu\mathrm{m}$, there is a brief remark on the clustering of
particles by capillary attraction in the absence of the electric
field: In these experiments the particles were in close contact in the
clustered states (with a mean interparticle separation, as inferred
from the photographs, of $\meanL \lesssim 2$ (in units of $R$)),
corresponding to states in the solid phase.  The lack of information
on the pre-clustered mean particle density renders our results in the
fluid phase of limited use for the comparison with the experimental
observations.  Nevertheless, assuming a hard--disk equation of state
for the particles in the absence of the electric field,
Fig.~\ref{fig:Kcharged}(a) indicates the formation of a cluster for
the range of particle sizes employed in this experiment, with collapse
times spanning several orders of magnitude (see
Fig.~\ref{fig:jeansT}(a)). 
The only quantitative experimental studies of the clustering
instability we are aware of which do provide data amenable to
comparison with our calculations can be found in
\rcites{HMZ94,VAKZ01}. According to the interpretation of those
authors, temperature inhomogeneities at the interface set the
particles in tomotion, the clustering of which is initially driven by
capillary forces; further restructuring of the emerging formations
inside the clusters involves short--ranged forces. This latter feature
is beyond the scope of our model and we are not in a position to judge
the influence of the temperature inhomogeneities.  We simply note
that, for buoyant particles of size $R \approx 35\,\mu\mathrm{m}$
spread with an initial areal density corresponding to $\meanL \approx
10$ \cite{VAKZ01}, Fig.~\ref{fig:Kcharged} predicts indeed
capillary--induced clustering and Fig.~\ref{fig:jeansT}(a) yields
Jeans' time of the order of a few minutes, in good agreement with the
reported characteristic times (see Fig.~3 of \rcite{VAKZ01}).
In summary, according to our estimates it
seems to be possible to perform experiments within an appropriate
range of controllable physical parameters which would promote the
occurrence of the capillary--induced instability in a variety of
conditions and which would allow the systematic study of its dynamical
evolution.

The tempting question arises whether the so far unexplained
interparticle attraction and the ensuing micron--sized clusters we
referred to in Sec.~\ref{sec:intro} and which are reported by various
groups can be understood within the physical picture we have
presented: a cluster would be held together against repulsion and
thermal diffusion by the collective capillary attraction and it would
be the final, equilibrium state of a capillary--induced collapse. The
answer is negative. First, there is a dynamical counterargument based
on our theoretical finding that a capillary--induced clustering would
require the almost simultaneous collapse of spontaneous density
fluctuations of all sizes (see \eq{eq:Tcoll2}): shortly after the
formation of the actually observed micrometer--sized clusters, there
should also arise many other clusters of larger sizes and all
particles would eventually gather in a single, large cluster. Such a
phenomenon has not been reported. Secondly, there is a static
counterargument based on the qualitative reasoning put forward in
Subsec.~\ref{sec:qualitative}, which states that the equilibrium size
of the cluster is of the order of Jeans' length associated with the
cluster density: as follows from Fig.~\ref{fig:Kcharged}, there is no
range of realistic values of the parameters for which Jeans' length is
of the order of the observed cluster size, typically tens of
micrometers at most, i.e., less than $10^{-2}\,\mathrm{mm}$. In other
words, the observed clusters consist of too few particles in order to
be able to build up a collective capillary attraction of relevant
strength.

One has also observed colloidal crystals spanning the whole system,
which has a typical size $L$ in the millimeter or centimeter range,
i.e., comparable with the capillary length $\lambda$ or somewhat
larger (see, e.g., \rcite{SCLN04,ASJN08}).
One could try to explain this particle distribution as a very large
cluster, self--confined by its own capillary attraction. Within our
model, this would correspond to a clustered state characterized by a
cluster Jeans' length $L_J$ of the order of $\lambda$ (see
Subsec.~\ref{sec:qualitative}).  Although our results pertain to the
fluid phase of the 2D colloid, one could nevertheless use them as
rough estimates for the solid phase, given the relatively weak
squared--root dependence of Jeans' length on the compressibility (see
\eq{eq:JeansK}). Thus, Figs.~\ref{fig:Kcharged} and
\ref{fig:Kelectric} provide evidence that the condition $L_J \sim
\lambda$ could be easily satisfied. Actually, \rcite{ASJN08} mentions
briefly the interpretation of the occurrence of large clusters as
being due to capillary attraction.

Recently, \citet{Perg09} has correctly pointed out the enhancement of
the pairwise capillary attraction due to a collective effect involving
many particles. His work addresses only the equilibrium configuration
of clusters of a size much larger than the capillary length and in the
solid phase. His somewhat involved calculations can be put in our
present context as follows: The solid phase is described effectively
by the equation of state of a harmonic solid of soft spheres, which
for the repulsive potential given by \eq{eq:urep} corresponds to a
pressure $p(\varrho) \propto \varrho^{1+(n/2)}$ at high densities
\cite{Spee03}. Integration of the equilibrium condition expressed by
\eq{eq:equil} yields $\varrho \propto (U - U_0)^{2/n}$, where $U_0$ is
an integration constant. Inserting this result into \eq{eq:meanU} one
recovers Eq.~(39) of \rcite{Perg09}, which is his central result
(notice that the exponent $n$ in our notation (see \eq{eq:urep})
corresponds to $n-1$ in Pergamenshchik's notation (see Eq.~(31) in
\rcite{Perg09})). These considerations show that the approach used in
\rcite{Perg09} is contained in the theoretical framework presented
here.

However, in view of the discussion given above his claims that such a
collective effect explains the clusters observed so far in experiments
appear to be unsubstantiated.
First, the analysis of an infinitely extended cluster misses the
explicit size--dependence of the capillary energy (see \eq{eq:ecap}),
which makes his results unapplicable to small clusters. Secondly, his
actual application of these results to experiments considers only
either an ideal gas or a gas of hard disks.  As we have shown in our
analysis, these two models do not give rise to significant differences
among them but both represent inadequate approximations for generic
conditions in typical experimental systems.  This difference can be
understood in terms of the non--capillary contribution to the energy
(see \eq{eq:Lcrit}): an ideal gas or a gas of hard disks contributes
kinetic energy only, which is of the order of the thermal energy $T$
per particle (in units of the Boltzmann constant), whereas for the
interparticle potential given by \eq{eq:urep}, the virial theorem
yields an energy per particle $e_\mathrm{short}/T = 1 + (2/n)
(p/(\varrho T) - 1)$, which according to Fig.~\ref{fig:eqState}(a) can
be at least $\approx 100 T$ in the solid phase.

The analogy of the model presented here with the 2D evolution of a
self--gravitating fluid is not complete: the capillary attraction is,
unlike gravity, screened beyond the capillary length, and the temporal
evolution is ruled by an overdamped dynamics (see \eq{eq:vel}, amended
in general with the effect of hydrodynamic interactions at
sufficiently high densities) rather than by the inertial, Newtonian
dynamics for gravitating particles. This poses the question as to
which extent the gravitational phenomenology can be reproduced by
colloids at a fluid interface.  Our study provides a partial answer,
in that it demonstrates the existence of a clustering instability
which is analogous to Jeans' instability. But this is still far from
being a complete and systematic comparison, which could provide the
tempting picture of the feasibility to study ``cosmology in a Petri
dish''. In this context, it would be interesting to investigate the
equilibrium configuration of clusters and their stability beyond the
simple qualitative analysis we have presented in
Subsec.~\ref{sec:qualitative}, i.e., by solving Eqs.~(\ref{eq:meanU})
and (\ref{eq:equil}). Such a study would be complementary to the
analysis of the dynamical aspects we have addressed here. The form of
the capillary attraction also leads to a possible analogy with
two--dimensional vortices, the similarity of which with a
self--gravitating system is also well known.  However, how deep and
useful this latter analogy can be is still a matter of study (see,
e.g., \rcite{Chav02a} and references therein).

\section{Conclusion}
\label{sec:conclusion}

We have presented a mean--field model for the evolution
of the density of colloidal particles at a fluid interface driven by
its own capillary attraction. In spite of the weakness of the
capillary interaction at the mean distance between particles, its
non-integrable character at submillimeter length scales enhances its
effect on the evolution of collective modes. We have demonstrated that
if the characteristic \textrm{Jeans' length} (see \eq{eq:JeansK}) of a
homogeneous distribution is sufficiently small (see
Fig.~\ref{fig:phases}) the system can be unstable with respect to
long--wavelength density perturbations under the action of capillary
attraction (see Fig.~\ref{fig:dispersion}). Beyond this linear
stability analysis, we have also solved the nonlinear evolution of
radially symmetric density perturbations (see Fig.~\ref{fig:shells})
in the so-called \textrm{cold--collapse approximation} (within which
the dynamics is driven by capillary attraction only) (see
Figs.~\ref{fig:shellR} -- \ref{fig:THprofile}), which predicts a
typical time of collapse of the order of Jeans' time (see
Eqs.~(\ref{eq:JeansK}) and (\ref{eq:Tcoll})). By computing Jeans'
length and time for presently accessible experimental setups we obtain
clear predictions about the range of parameter values within which the
instability could be observed. Jeans' time (see Fig.~\ref{fig:jeansT})
depends on the strength of the capillary monopole. To this end we have
considered the monopole to be induced either by buoyancy or by an
external electric field. Jeans' length depends additionally on the
equation of state of the 2D gas via its compressibility. In this
context we have studied an ideal gas (see Fig.~\ref{fig:Kcharged}(a))
and systems with a dipolar interparticle repulsion (see \eq{eq:urep}
for $n=3$) induced either by electric charges on the particles (see
Fig.~\ref{fig:Kcharged}(b)), an external electric field (see
Fig.~\ref{fig:Kelectric}), or an external magnetic field (see
Fig.~\ref{fig:Kmagnetic}). In most experiments performed so far, the
physical parameters lie in the region of stability, but they appear to
be easily tunable into the instability regime. The relatively weak
dependence of Jeans' length on the equation of state (via the square
root of the compressibility, see \eq{eq:JeansK}), renders the
capillary monopole to be a possibly more convenient parameter for
tuning Jeans' length and time. Experiments with 2D colloids exposed to
an external electric field seem particularly promising in this
respect, because the field provides a simple way of controlling the
vertical pull on the particles generating in turn the mediating
interfacial deformation. Similarly, a vertical force on
superparamagnetic particles in an external magnetic field could be
created and controlled via gradients of the magnetic field which pull
on the induced magnetic moments.

\acknowledgments

A.D.~acknowledges support by the Ministerio de Educaci\'on y Ciencia
(Spain) through Grant Number FIS2008-01339 (partially financed by
FEDER funds). M.O.~thanks the German Research Foundation (DFG) for
financial support through the Collaborative Research Centre (SFB--TR6)
``Colloids in External Fields", {project N01}.

\appendix

\section{Functional formulation}
\label{sec:functional}

The mathematical model defined by Eqs.~(\ref{eq:meanU}) and
(\ref{eq:diff}) can be reformulated in terms of the functional ${\cal
  F}[\varrho, U] = {\cal F}_{\rm cap} + {\cal F}_{\rm gas} + {\cal
  F}_{\rm inter}$, consisting of three contributions. The first one is
related to the capillary deformation in the small--deformation limit:
\begin{equation}
  {\cal F}_{\rm cap} := \frac{1}{2}\gamma \int dA\; \left[ |\nabla U|^2
    + \left(\frac{U}{\lambda}\right)^2 \right] .
\end{equation}
The second term is the free energy functional (within local
approximations) of the 2D gas of particles,
\begin{equation}
  {\cal F}_{\rm gas} := \int dA\; \mathsf{f_{gas}}(T, \varrho) ,
  \qquad
  \mathsf{f_{gas}}(T, \varrho) = T \varrho [\ln (\Lambda^2 \varrho) -1 ] + 
  \mathsf{f_{ex}}(T, \varrho) ,
\end{equation}
where the free energy density $\mathsf{f_{gas}}(T, \varrho)$ is the
sum of the ideal gas contribution ($\Lambda$ is de Broglie's thermal
length) and the excess free energy $\mathsf{f_{ex}}$ due to the
repulsive short--ranged forces. Finally, the third term represents the
interaction between the particles and the interfacial deformation:
\begin{equation}
  \label{eq:Finter}
  {\cal F}_{\rm inter} := - f \int dA\; \varrho\, U .
\end{equation}
The mean--field equation (\ref{eq:meanU}) for the interfacial
deformation follows from the extremal condition
\begin{equation}
  \frac{\delta{\cal F}}{\delta U(\br)} = 0 ,
\end{equation}
while the diffusion equation (\ref{eq:diff}) can be expressed in terms
of a relaxation--type dynamics,
\begin{equation}
  \label{eq:relaxDyn}
  \frac{\partial \varrho}{\partial t} = 
  \Gamma \nabla\cdot\left[\varrho \nabla 
    \frac{\delta{\cal F}}{\delta \varrho(\br)} 
    \right] ,
\end{equation}
upon using the thermodynamical identity
\begin{equation}
  \label{eq:chemicalPot}
  p(\varrho) = \varrho^2 \frac{\partial}{\partial \varrho}
    \left(\frac{\mathsf{f_{gas}}(T, \varrho)}{\varrho}\right)_T
    \qquad\Rightarrow\qquad
    \nabla p = \varrho \nabla \left(
      \frac{\partial \mathsf{f_{gas}}}{\partial\varrho}\right) 
    \quad \textrm{at constant } T .
\end{equation}
In principle, ${\cal F}$ can be viewed as an effective functional
reformulation of the problem, although it can be associated with an
actual free--energy functional for the physical system. In this case,
the mean--field approximation enters via the simplified form of
\eq{eq:Finter}: a complete description of the particle--interface
interaction should take into account the finite size of the particles
(rather than a point capillary monopole), their shape, the
corresponding surface energies (see, e.g., the free--energy functional
introduced in \rcite{ODD05a}), and a thermal noise contribution to
\eq{eq:relaxDyn}.
A procedure for incorporating the hydrodynamic interactions into the
functional formulation has been proposed recently in \rcite{ReLo09}.

\section{Equation of state from numerical simulations}
\label{sec:MC}

We investigate a two--dimensional fluid governed by the pair potential
given by the power law in Eq.~(\ref{eq:urep}) with $n=3$.
The equation of state of such a fluid depends only on the
dimensionless parameter $\zeta^2 \rho$ \cite{HGJ71}. Thus it
is advantageous to introduce the new length scale
$\tilde\sigma=1/\sqrt{\rho}$ such that
\begin{equation}
  \label{eq:MCpotential}
 \frac{v_\mathrm{rep}(d)}{T} = (\zeta^2 \rho)^{3/2} 
 \left( \frac{\tilde \sigma}{d} \right)^3\;.
\end{equation}
In the simulation one fixes $\tilde\sigma$ as the unit of length and
varies the prefactor $(\zeta^2 \rho)^{3/2}$. The pressure can be
determined from the virial expression
\begin{equation}
  p = \rho T + \frac{1}{2 A} \langle \sum_{i<j}  
  {\bf f}({\bf r}_i-{\bf r}_j) \cdot ({\bf r}_i-{\bf r}_j) \rangle
\end{equation}
where $A$ is the area of the simulation box and ${\bf f}= -\nabla
v_\mathrm{rep}$ is the interparticle force. In our case
\begin{equation}
 \sum_{i<j}  {\bf f}({\bf r}_i-{\bf r}_j)
  \cdot ({\bf r}_i-{\bf r}_j) = 3  \sum_{i<j} v_\mathrm{rep}({\bf r}_i-{\bf r}_j)\;,
\end{equation}
so that the pressure is related to the excess internal energy $U^{\rm
  ex}$ of the fluid according to
\begin{equation}
  p = \rho T + \frac{3}{2} \frac{U^{\rm ex}}{A}\;.
\end{equation}
The simulations have been carried out with $N=400$ particles in a
square simulation box with reduced side length $L=20$ (in units of
$\tilde\sigma$).
Periodic boundary conditions were applied. The internal energy of a
given configuration of the $N$ particles was obtained by summing over
$(2l+1)\times(2l+1)$ boxes (i.e., the central simulation box plus
image boxes around it) because the pair potential decays slowly (we
chose $l=3$). $U^{\rm ex}$ was determined as an average over $10^4$
sweeps (one sweep corresponds to attempted moves for every particle in
the box). The instantaneous internal energy was determined for each
sweep once.

\section{Radially symmetric cold collapse}
\label{sec:collapse}

\begin{figure}[h]
  \centering\includegraphics[width=.25\textwidth]{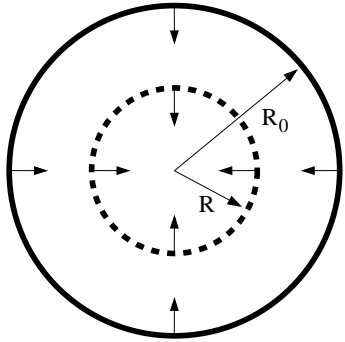}
  \caption{Evolution of a radially symmetric enhanced density. An
    infinitesimally thin ring of particles of initial radius $R_0$
    collapses to a ring of radius $R(t) < R_0$ at a time $t>0$.  By
    construction the ring follows the motion of its particles and it
    can be identified by its Lagrangian radial coordinate $R_0$ (and
    $R(t)$ is the Eulerian radial coordinate of the ring).  Therefore,
    the number of particles inside the disk encircled by the
    ring is constant if different rings do not cross (one cannot
    exclude a priori that a ring is overtaken by another one because
    different rings may collapse with different speeds).}
  \label{fig:shells}
\end{figure}

\subsection{General dynamics of radially symmetric cold collapse}

\begin{figure}[h]
  \centering\includegraphics[width=.5\textwidth]{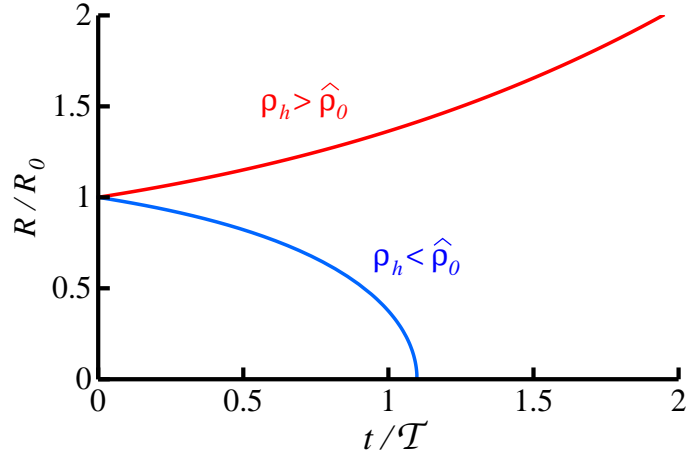}
  \caption{\textcolor{blue}{(color online)} Evolution in time of the
    radius $R(t; R_0)$ of a circular ring during cold collapse
    (\eq{eq:radiusR}). Two qualitatively different cases (expansion
    and collapse) can be distinguished depending on the initial
    average density $\hat{\varrho}_0$ in the area encircled by the
    ring compared to the background density $\varrho_h$. The curves
    correspond to $\hat{\varrho}_0/\varrho_h = 0.5$ for expansion and
    $\hat{\varrho}_0/\varrho_h = 1.5$ for collapse.}
  \label{fig:shellR}
\end{figure}

A homogeneous configuration is characterized by a density $\varrho_h$
and a mean--field interfacial deformation $U_h = f \lambda^2 \varrho_h
/ \gamma$ given by \eq{eq:meanU}. We study the evolution of a radially
symmetric density perturbation so that
\begin{equation}
  \varrho(\br, t) = \varrho_h + \delta \varrho (r, t) .
\end{equation}
The perturbation $\delta U$ of the interfacial deformation is given by
the solution of \eq{eq:linU}. (Note that this latter equation is not a
linearized approximation requiring $\delta\varrho$ and $\delta U$ to
be small, because \eq{eq:meanU} \textit{is} linear; the nonlinearity
of the dynamics enters via \eq{eq:diff}.)  We consider a sufficiently
localized density perturbation, i.e., $\delta \varrho$ vanishes
sufficiently fast at infinity, so that the solution in the limit of
large capillary length $\lambda$ follows immediately from the
gravitational analogy: the capillary force on a particle is (compare
\eq{eq:Vcap_mono})
\begin{equation}
  \label{eq:deltaU}
  f \nabla (\delta U) =
  - \frac{f^2 \, \delta N(r,t)}{2\pi\gamma r} \be_r ,
  \qquad (\lambda \to \infty)
\end{equation}
where
\begin{equation}
  \delta N(r,t) := 2\pi \int_0^r ds\, s\, \delta\varrho(s,t)
\end{equation}
is the excess number of particles in a disk of radius $r$ concentric
with the perturbation. The density field follows from solving
Eqs.~(\ref{eq:cont})--(\ref{eq:diff}), which in the cold collapse
approximation, i.e., after neglecting any force other than the
capillary attraction, reduce to (see \eq{eq:deltaU})
\begin{equation}
  \label{eq:cont2}
  \frac{\partial (\delta\varrho)}{\partial t} = 
  - \nabla\cdot(\bv \varrho) ,
\end{equation}
\begin{equation}
  \label{eq:coldV}
  \bv = \Gamma f \nabla (\delta U) .
\end{equation}
An exact solution is possible by resorting to a description of the
flow in terms of Lagrangian coordinates. Consider the thin ring of
radius $R(t; R_0)$ formed by the particles which were initially
($t=0$) at a distance $R_0$ from the center, i.e., the ring is defined
through its Lagrangian radial coordinate $R_0$ (see
Fig.~\ref{fig:shells}).  Equation~(\ref{eq:cont2}) implies that the
number of particles encircled by the ring is conserved\footnote{In
  this appendix, a subindex $0$ denotes evaluation at the initial time
  $t=0$.}:
\begin{equation}
  \label{eq:Nconserved}
  N(R, t) := 2\pi \int_0^R ds\, s\, \varrho(s,t) = N_0(R_0) = 
  2\pi \int_0^{R_0} ds\, s\, \varrho_0(s) ,
\end{equation}
where $N_0(R_0)$ is the initial number of particles encircled by the
ring of radius $R_0$. (Since, at fixed $t$, $N(R,t)$ is a monotonous
function of $R$, \eq{eq:Nconserved} provides a relation $R_0 = R_0 (t;
R)$ which can be inverted, $R = R(t; R_0)$, with the geometrical
meaning ``radius of the ring at time $t$ which had a radius $R_0$ at
time $t=0$''.) This expression allows an explicit computation of the
density field,
\begin{equation}
  \label{eq:evolvedProfile}
  \varrho (R, t) = \frac{1}{2\pi R} \left.
    \frac{\partial N}{\partial R}\right|_t = 
  \frac{1}{2\pi R} \frac{d N_0}{d R_0} 
  \left. \frac{\partial R_0}{\partial R}\right|_t =
  \frac{R_0}{R} \varrho_0 (R_0) \left.
    \frac{\partial R_0}{\partial R}\right|_t .
\end{equation}
The evolution of the ring radius $R(t; R_0)$ is determined by the
flow velocity,
\begin{equation}
  \label{eq:inflow}
  \left.\frac{\partial R}{\partial t}\right|_{R_0} \be_r = \bv ,
\end{equation}
with $\bv$ given by \eq{eq:coldV}. By inserting the
solution~(\ref{eq:deltaU}) and making use of particle conservation,
$\delta N (R, t) = N(R, t) - \pi R^2 \varrho_h = N_0 (R_0) - \pi R^2
\varrho_h$ (see \eq{eq:Nconserved}), one finally arrives at an
ordinary, first--order differential equation for $R(t; R_0)$:
\begin{equation}
  \left.\frac{\partial R}{\partial t}\right|_{R_0}
  = \left. \Gamma f \be_r\cdot\nabla (\delta U)\right|_{r=R} = 
  - \frac{f^2 \Gamma}{2\pi\gamma R} [ N_0(R_0) - \pi R^2 \varrho_h] .
\end{equation}
In terms of the characteristic time $\jeansT = \gamma/(\Gamma f^2
\varrho_h)$ of the homogeneous background configuration (see
\eq{eq:JeansK}) one can write
\begin{equation}
  \jeansT \left.\frac{\partial}{\partial t} 
    \left(\frac{R}{R_0}\right)^2\right|_{R_0} = 
  \left(\frac{R}{R_0}\right)^2 - \frac{\hat\varrho_0}{\varrho_h} ,
\end{equation}
where
\begin{equation}
  \label{eq:averageRho}
  \hat\varrho_0 (R_0) := \frac{N_0(R_0)}{\pi R_0^2}
\end{equation}
is the average initial density of the disk of radius $R_0$. The
solution of this equation with the boundary condition $R(t=0; R_0) =
R_0$ reads
\begin{equation}
  \label{eq:radiusR}
  R(t; R_0) = R_0 \sqrt{\frac{\hat\varrho_0}{\varrho_h} + 
    \left( 1 - \frac{\hat\varrho_0}{\varrho_h} \right) 
    {\rm e}^{t/\jeansT}} .
\end{equation}
Figure~\ref{fig:shellR} plots the evolution $R(t)$ for the two
possible, qualitatively different cases: If $\varrho_h >
\hat\varrho_0$ the ring expands. (When $R$ reaches $\lambda$ the model
ceases to be valid because then one must take into account the
screening of the capillary interaction neglected in \eq{eq:deltaU}.)
In the other case, $\varrho_h < \hat\varrho_0$, the ring contracts and
$R=0$ is reached at a time
\begin{equation}
  \label{eq:Tcoll}
  \jeansT_{\rm coll} = - \jeansT \ln \left( 1 - 
    \frac{\varrho_h}{\hat\varrho_0} \right) .
\end{equation}
Obviously, true collapse is prevented by a finite compressibility
which has been neglected in the above calculations (see \eq{eq:coldV})
and a spatially extended equilibrium configuration will emerge.
However, the unambiguos identification of a ring by its Lagrangian
radius $R_0$ would no longer hold if two different rings would
coincide before the collapse at the center (a phenomenon customarily
called {\em shell crossing} in 3D in the cosmological literature). If
this would happen, a density singularity would appear, maybe
regularized by the repulsive forces present. It can be shown that a
sufficient and necessary condition to avoid ring crossing is that the
time of collapse $\jeansT_{\rm coll} (R_0)$ of the ring characterized
by its initial radius $R_0$ grows with this radius. In view of
\eq{eq:Tcoll}, this is equivalent to the condition
\begin{equation}
  \frac{d\hat\varrho_0}{d R_0} \leq 0 ,
\end{equation}
i.e., the initial perturbation inside a disk looks more and more
rarefied as the disk radius is taken larger and larger.

\subsection{Example of cold collapse for a steplike overdensity}

\begin{figure}[h]
  \centering\includegraphics[width=.5\textwidth]{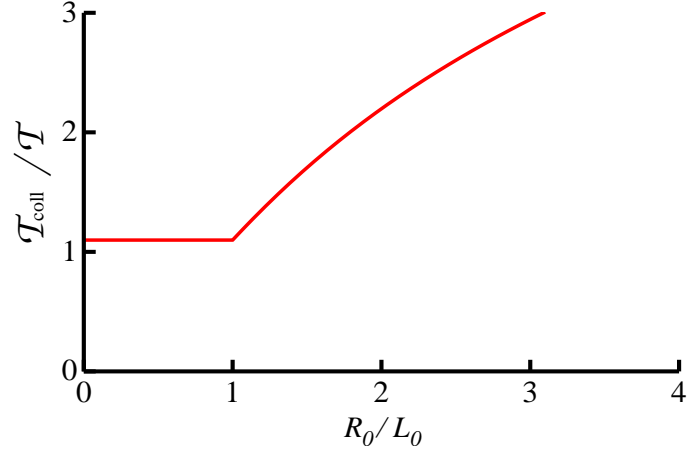}
  \caption{\textcolor{blue}{(color online)} Dependence of the time of
    collapse on the ring radius for a top--hat enhanced density
    distribution. The plot corresponds to the case $\delta \varrho_0 =
    0.5 \varrho_h$ (see \eq{eq:TcollTH}). The time scale is set by
    Jeans' time $\jeansT$ (\eq{eq:JeansK}).}
  \label{fig:Tcoll}
\end{figure}

\begin{figure}[h]
  \centering\includegraphics[width=.5\textwidth]{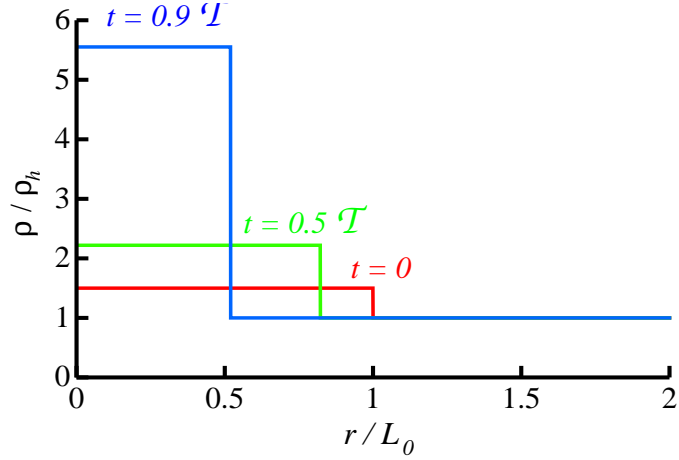}
  \caption{\textcolor{blue}{(color online)} Time evolution of the
    density profile of an initial top--hat enhanced density
    distribution (\eq{eq:evolutionTHRho}). The plot corresponds to
    $\delta \varrho_0 = 0.5 \varrho_h$. The time scale is set by
    Jeans' time $\jeansT$ (\eq{eq:JeansK}).}
    \label{fig:THprofile}
\end{figure}

As a simple application of these results we consider a top--hat
enhanced density, described by the initial profile 
\begin{equation}
  \label{eq:THRho}
  \varrho_0(r) = 
  \varrho_h + \delta \varrho_0 \, \Theta(L_0 - r) =
  \left\{
    \begin{array}[c]{cc}
      \varrho_h + \delta\varrho_0 , & r < L_0 \\
      \varrho_h , & L_0 < r
    \end{array} \right.
  \qquad (\delta\varrho_0 > 0) ,
\end{equation}
where $\Theta(x)$ is Heavyside's step function. We assume that the
initial radius fulfills $K^{-1} \ll L_0 \ll \lambda$, so that the
simplifying assumptions of the previous calculations hold. For this
profile, one has (\eq{eq:averageRho})
\begin{equation}
  \label{eq:averageTHRho}
  \hat\varrho_0 (R_0) = \left\{
    \begin{array}[c]{cc}
      \varrho_h + \delta\varrho_0 , & R_0 \leq L_0 \\
      & \\
      \displaystyle
      \varrho_h + \left(\frac{L_0}{R_0}\right)^2 \delta\varrho_0 , 
      & L_0 < R_0 ,
    \end{array} \right.
\end{equation}
so that $\varrho_h < \hat\varrho_0$ and, according to
Fig.~\ref{fig:shellR}, there is a collapse for all $L_0 > 0$. The time
of collapse (see \eq{eq:Tcoll})
\begin{equation}
  \label{eq:TcollTH}
  \jeansT_{\rm coll} (R_0) = \jeansT \ln \left\{ 1 + \frac{\varrho_h}{\delta\varrho_0} 
    \left[ \Theta (L_0 - R_0)  + 
      \left(\frac{R_0}{L_0} \right)^2 \Theta (R_0 - L_0) 
    \right] \right\}
\end{equation}
is plotted in Fig.~\ref{fig:Tcoll}: the whole interior of the
overdensity ($R_0 < L_0$) collapses simultaneously, i.e.,
$\jeansT_\mathrm{coll}$ is independent of $R_0$. The evolution of the
density profile $\varrho(r, t)$ follows from
Eqs.~(\ref{eq:evolvedProfile}) and (\ref{eq:THRho}): by inserting
\eq{eq:averageTHRho} into \eq{eq:radiusR} one obtains upon
differentiation
\begin{equation}
  \left.\frac{R}{R_0} \frac{\partial R}{\partial R_0} \right|_t= 
  \left\{ 
    \begin{array}[c]{cc}
      (R/R_0)^2 , & R_0 < L_0 \\
      1 , & L_0 < R_0
    \end{array}\right.
\end{equation}
so that the density profile preserves the top--hat shape during the
evolution (see Fig.~\ref{fig:THprofile}):
\begin{equation}
  \label{eq:evolutionTHRho}
  \varrho (r, t) = \left\{
    \begin{array}[c]{cc}
      \displaystyle
      \left[\frac{L_0}{L(t)}\right]^2 (\varrho_h + \delta\varrho_0) ,
      & r < L(t) \\
      & \\
      \varrho_h , & L(t) < r
    \end{array} \right.
\end{equation}
where
\begin{equation}
  L(t) = L_0 \sqrt{\frac{\varrho_h + \delta\varrho_0}{\varrho_h}
    \left[ 1 - {\rm e}^{(t-\jeansT_{\rm coll})/\jeansT} \right] }
\end{equation}
is the time--dependent radius of the region with enhanced density,
i.e., \eq{eq:radiusR} evaluated for the particular case $R_0=L_0$ by
using Eqs.~(\ref{eq:Tcoll}) and (\ref{eq:averageTHRho}). Consistently
with this geometrical meaning, $L(t)$ vanishes at the time of collapse
$t = \jeansT_\mathrm{coll} (R_0 = L_0)$ of this region, as given by
\eq{eq:TcollTH}. (Upon deriving \eq{eq:evolutionTHRho} one notices
that the region with enhanced density can be characterized by the
inequality $R_0 < L_0$ or equivalently by $r < L(t)$, since $r = R(t;
R_0)$.) The sharp jump in the density profile is of course an
idealization in the limit of infinite compressibility. The jump would
be actually smoothed out by the pressure forces, but on a length scale
much smaller than the spatial extension of the density enhancement.


As the perturbation is compressed, in the 2D fluid the pressure
increases until it is able to counterbalance the total capillary force
and stop the collapse. One can derive a simple estimate for the
smallest size $L_\mathrm{min}$ of the compressed cluster as follows:
at the maximum density $\varrho_{\rm max} \gg \varrho_h$,
corresponding to a cluster radius $L_{\rm min} \ll L_0 \ll \lambda$, a
total \textit{p}ressure force\footnote{Note that in 2D pressure is a
  force per unit length.}  of the order of $F_{\rm p} \sim 2\pi L_{\rm
  min} p(\varrho_{\rm max})$ opposes the total compressing capillary
force, which is of the order of $F_{\rm cap} \sim (f N)^2/(2\pi\gamma
L_{\rm min})$. (The cluster consists of $\approx N$ particles and the
capillary force on each of them is given by \eq{eq:deltaU} with
$\delta N = \pi L_{\rm min}^2 (\varrho_{\rm max} - \varrho_h) \approx
N$ due to $\varrho_{\rm max} \gg \varrho_h$.) Therefore, in
equilibrium $F_{\rm p} \sim F_{\rm cap}$, giving the relationship
\begin{equation}
  p(\varrho_{\rm max}) \sim 
  \frac{(f L_{\rm min} \varrho_{\rm max})^2}{4\gamma} .
\end{equation}
Upon introducing Jeans' length $K_{\rm cluster}^{-1}$ of the cluster,
i.e., \eq{eq:JeansK} evaluated at the cluster average density
$\varrho_{\rm max}$, this can be rewritten as
\begin{equation}
  \label{eq:equilibriumL}
  (K_{\rm cluster} L_{\rm min})^2 \sim 
  4 p(\varrho_{\rm max}) \kappa(\varrho_{\rm max}) .
\end{equation}
For polytropic equations of state, $p(\varrho) \propto \varrho^z$, or
more generally for a simple fluid far from any phase transition, the
right hand side of this expression is typically of the order of unity,
so that the final equilibrium size of the cluster will be comparable
to its Jeans' length. This rough estimate agrees with the results
obtained from the qualitative reasoning given in
Subsec.~\ref{sec:qualitative}.


\end{document}